\documentclass[12pt,pra,aps,preprint,superscriptaddress,showpacs,epsfig,floatfix]{revtex4-1}
\usepackage{color}
\usepackage{graphicx,amssymb,epsfig,epsf,amsmath,fontenc}
\def\beq{\begin{equation}}
\def\eeq{\end{equation}}
\def\beqn{\begin{eqnarray}}
\def\eeqn{\end{eqnarray}}

\newcounter{saveeqn}

\def\C {{\bf C}}

\def\brho {\mbox{\boldmath $\rho$}}

\date{\today}%\vspace{-5ex}}
\begin{document}
\title{Many-body quantum dynamics of an asymmetric bosonic Josephson junction}

\author{Sudip Kumar Haldar}
%\email[Email:]{shaldar@campus.haifa.ac.il}
\author{Ofir E. Alon}
%\email[Email:]{ofir@research.haifa.ac.il}
\affiliation{Department of Mathematics, University of Haifa, Haifa 3498838, Israel.}
\affiliation{Haifa Research Center for Theoretical Physics and Astrophysics,University of Haifa, Haifa 3498838, Israel.}

\begin{abstract}
The out-of-equilibrium quantum dynamics of an interacting Bose gas trapped in a one-dimensional asymmetric double-well potential
is studied by solving the many-body Schr\"odinger equation numerically accurately. We examine how the loss of symmetry of the confining 
trap affects the macroscopic quantum tunneling dynamics of the system between the two wells. In an asymmetric double well, the two wells 
are not equivalent anymore - the left well is deeper than the right one. Accordingly, we analyze the dynamics by initially preparing 
the condensate in both the left and the right well. The dynamics of the system is characterized by the time evolution of a few physical 
quantities of increasing many-body complexity, namely, the survival probability, depletion and fragmentation, and the many-particle position
and momentum variances. In particular, we have examined the frequencies and amplitudes of the oscillations of the survival probabilities, the 
time scale for the development of fragmentation and its degree, and the growth and oscillatory behavior of the many-particle position and 
momentum variances. There is an overall suppression of the oscillations of the survival probabilities in an asymmetric double well. However, 
depending on whether the condensate is initially prepared in the left or right well, the repulsive inter-atomic interactions affect 
the survival probabilities differently. For a sufficiently strong repulsive interaction, the system is found to become fragmented. The 
degree of fragmentation depends both on the asymmetry of the trap and the initial well in which the condensate is prepared in a non-trivial 
manner. Overall, the many-particle position and momentum variances bear the prominent signatures of the density oscillations of the system in the 
asymmetric double well as well as a breathing-mode oscillation. Finally, a universality of fragmentation for systems made of different numbers 
of particles but the same interaction parameter is also found. The phenomenon is robust despite the asymmetry of the junction and admits a 
macroscopically-large fragmented condensate characterized by a diverging many-particle position variance. This is as far as one can get from the 
dynamics of the density in the junction.
\end{abstract}

\maketitle
\section{Introduction}

The dynamics of ultra-cold quantum gases has attracted a lot of interest since the experimental observations of Bose-Einstein condensation (BEC)~\cite{ex1,ex2,ex3}. 
The advent of advanced trapping techniques and controlling of inter-particle interactions has 
made it possible to experimentally study several problems which were elusive until recently. This has opened a whole new research field of strongly 
correlated systems with potential applications in various fields such as quantum computing and quantum simulation of condensed-matter 
problems~\cite{Feynman,lewenstein2007,lewenstein2012,sowinski2010}. One such well-studied example 
is the system of a few interacting bosons in a double-well potential~\cite{Shenoy,Gati,Milburn, Jacek, Junpeng, menotti2001,meier2001,salgueiro2007,zollner2008,Carr2010,LeBlanc,simon2012,he2012,Gillet,liu2015,tylutki2017,dobrzyniecki2018,Dobrzyniecki}.

A symmetric double-well potential provides a paradigm model for many physical systems such as the bosonic Josephson junction (BJJ)~\cite{Gati2007}.
BJJ dynamics has been studied quite thoroughly both theoretically and 
experimentally~\cite{Shenoy, Gati, Milburn, Jacek, Junpeng,LeBlanc,Gillet,Gati2007, Shmuel, PRB2010, Levy, Raghavan, Ostrovskaya, Zhou, Lee, Ananikian,Ferrini,Schesnovich,Jia,Trujillo,Zibold, Spagnolli, Burchinati, Sakmann2009, Sakmann2010, Sakmann2014, Sudip2018}. 
Several features like Josephson oscillations~\cite{Shenoy, Gati, LeBlanc, Gillet, Levy, Zibold, Spagnolli, Burchinati}, 
collapse and revival cycles~\cite{Milburn}, self trapping (suppression of tunneling)~\cite{Gati,Milburn,Shenoy,Gillet,Levy,Zibold}, etc. have been predicted using a two-mode theory and later experimentally observed~\cite{Gati}. Recently, BJJ dynamics has also been studied by 
an in-principle numerically-exact many-body theory~\cite{Sakmann2009,Sakmann2010,Sakmann2014}.
In particular, fragmentation~\cite{Sakmann2014,MCHB} and the uncertainty product of the many-particle position
and momentum operators~\cite{Klaiman2016} have been studied by solving the many-body Schr\"odinger equation. 
Further, a universality of fragmentation in the sense that systems with different particle numbers $N$, 
keeping the interaction parameter $\Lambda=\lambda_0(N-1)$ fixed ($\lambda_0$ being the strength of interaction), fragment to the same value~\cite{Sakmann2014} has been predicted in 
the dynamics of interacting bosons in a symmetric double well for a sufficiently strong interaction. Also,  
the impact of the range of the interaction on the dynamics of a BJJ has been investigated recently~\cite{Sudip2018}.

Symmetry breaking is of fundamental interest in physics. Accordingly, an asymmetric double well is of particular interest. 
{Already a number of studies of the properties of an ultra-cold atomic system in an asymmetric double well trap and a few of its applications have been reported}~\cite{Hall,PRA2010,Hunn2013,PRA2014,Carvalho2015,PRL2016,PRA2016kim,PRA2016Paul,Cosme2017}.
For example, a novel sensor utilizing the 
adiabatic axial splitting of a BEC in an asymmetric double well has been reported~\cite{Hall}.
The ground state properties of spin-$1$ bosons~\cite{Carvalho2015} in an asymmetric double well has also been studied.
Also, the ground state properties and the corresponding transition between the Josephson and self-trapped regimes for an attractive BEC have been studied by the two-site Bose-Hubbard model~\cite{PRA2010}.
Moreover, the tunneling of a two-boson system~\cite{Hunn2013} and 
the interaction blockade for a few boson-system {with up to $N=3$ bosons}~\cite{Cosme2017} in an asymmetric double well have also been studied. 
A two-mode model has been constructed to study the dynamics of a BEC in an asymmetric 
double well and its phase-space properties are analyzed~\cite{PRA2014}. %Further, an out-of-equilibrium dynamics of a Bose-Hubbard dimer following a quench from a highly asymmetric to symmetric trap have been analyzed in~\cite{PRB2010}. 
{However, a systematic study of the dynamics of a many-particle bosonic system in an asymmetric double well for different interaction regimes using a numerically-exact many-body method is, to the best of our knowledge, yet to be reported. Such a method automatically includes all participating bands in the asymmetric double well. This allows us to describe the physics of the asymmetric BJJ both when it is fully condensed and when it becomes fragmented on an accurate many-body level.} 

Therefore in this work, we ask how the loss of symmetry in the double well trap may affect the many-body physics of BJJ dynamics for different strengths of interaction. Since BJJ dynamics involves macroscopic quantum tunneling, such studies are of general interest. 
Here we consider a short-range contact $\delta$ interaction of tunable strength $\lambda_0$ which is the popular model for inter-atomic 
interaction in ultra-cold atomic systems~\cite{rev1}. In this work, we examine the impact of different degrees of asymmetries, for different strengths of interactions, on the dynamics of BEC in an asymmetric double well following a trapping quench
from a single harmonic well to the asymmetric double well at time $t=0$. Moreover, in an asymmetric double well, the two wells are not equivalent anymore and for the kinds of asymmetric double wells considered in this work, the left well is lower than the right well. Accordingly, 
we study the non-equilibrium dynamics of the system following the trapping quench by preparing the BEC in the left and the right wells, and ascertain how the dynamics depends on the initial well. 

For our study, we numerically accurately solve the many-body Schr\"odinger equation~\cite{MCHB} and 
{characterize the dynamics of the system by the time evolution of a few physical quantities of varying degrees of complexity, both at the 
mean-field and the many-body level. We focus on the time evolution of the survival probability, depletion and fragmentation, and the variance of the many-particle position and momentum operators. We examine both the weakly-interacting system as well as that with a stronger interaction where the system becomes fragmented, and thereby explore how the many-body features develop in these quantities in different interaction regime. Therefore, in this work, our scope of investigation is far beyond that of Ref~\cite{Sakmann2014} where only the strong interaction case was considered. More importantly for the first time, the time evolution of the many-particle position and momentum variances in the junction are
discussed in this work.}
%and the dynamics of the system were not discussed at the mean-filed level and therefore the developments of the many-body effects
%with interaction were not discussed.}
%Further, we also characterize the physical quantities as well as the 
%strengths of interaction for which the dynamics of the system can be accurately described at the mean-field level.

The density oscillations of a BEC is found to be suppressed in an asymmetric double well. However, the repulsive inter-atomic interaction facilitates the tunneling between the two wells when the initial condensate is prepared in the left well. On the other hand, if the initial BEC is prepared in the right well, the repulsive interaction suppresses the oscillations further. For a stronger interaction, the BEC becomes fragmented and the degree of fragmentation is found to depend on the initial well. Further, a universality of the fragmentation dynamics is also observed, though again the degree of the universal fragmentation differs for the left and the right well. We also found prominent signatures of density oscillations as well as breathing-mode oscillations 
in the time evolution of the variances of the many-particle position and momentum operators. {Note that for the description of the breathing mode oscillations, one needs to take into account the coupling with higher energy bands and, therefore, it is beyond the scope of Bose-Hubbard dimer.}

This paper is organized as follows. In Section~\ref{Theory}, we introduce the quantities which will be used to characterize the dynamics and also the
in principle numerically-exact many-body method used to solve
the time-dependent many-body Schr\"odinger equation. In Section~\ref{results}, we present and
discuss our findings. Finally, we summarize and put our concluding remarks in Section~\ref{conclusions}. Numerical convergence is discussed in the Appendix.

\section{Theoretical framework}
\label{Theory}

In this section, we introduce the theoretical methods and quantities used in this work to explore the dynamics of an asymmetric bosonic Josephson junction. 

\subsection{System}

\begin{figure}
\begin{center}
\includegraphics[width=0.75\linewidth]{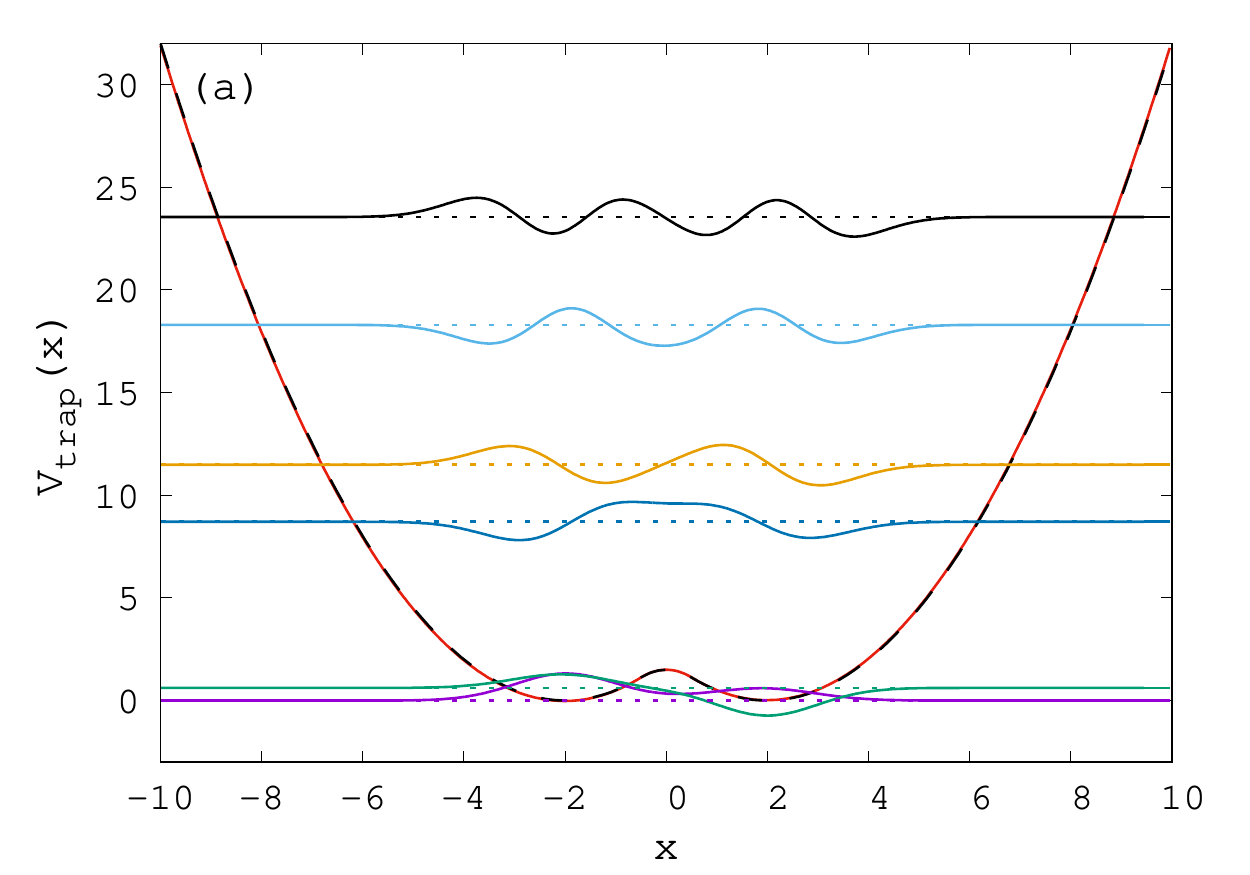}\\
\includegraphics[width=0.45\linewidth]{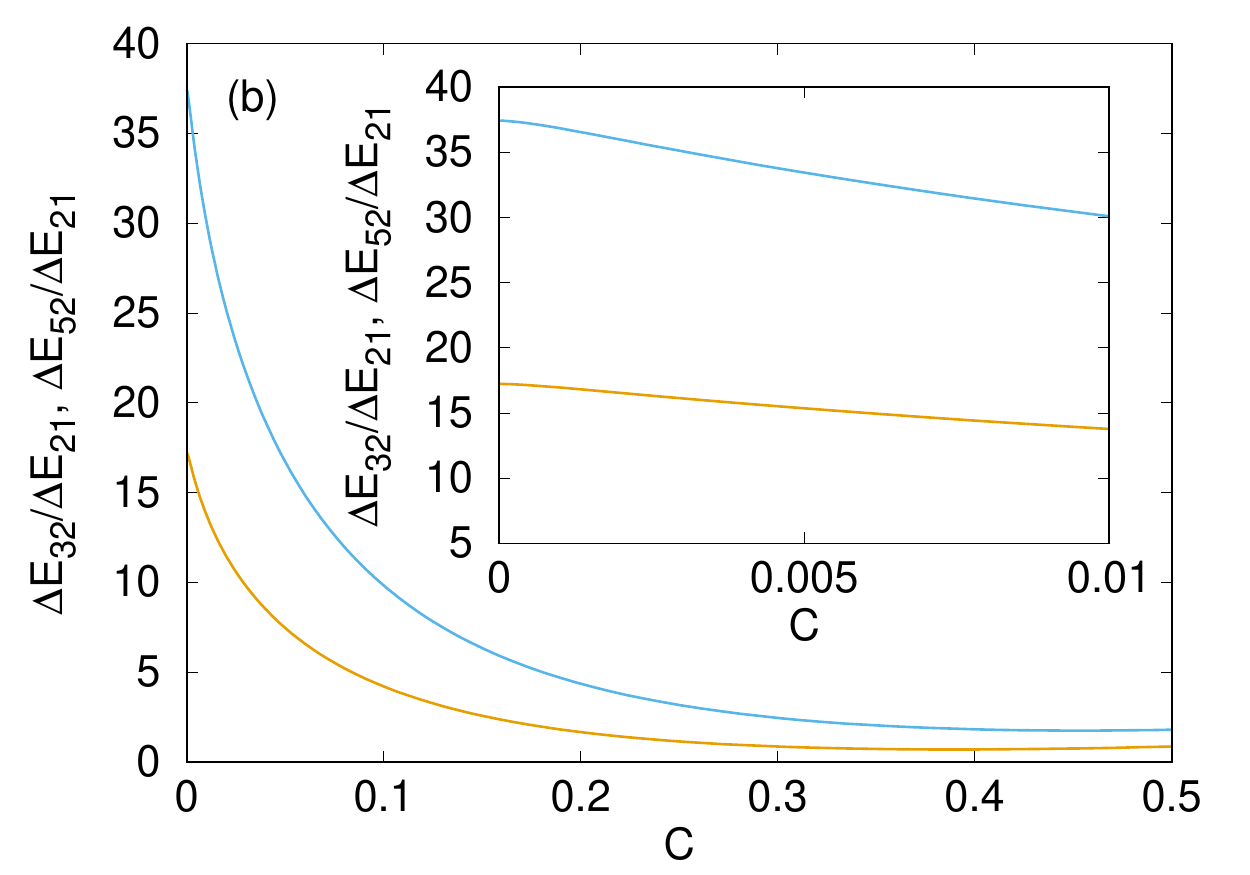}
\includegraphics[width=0.45\linewidth]{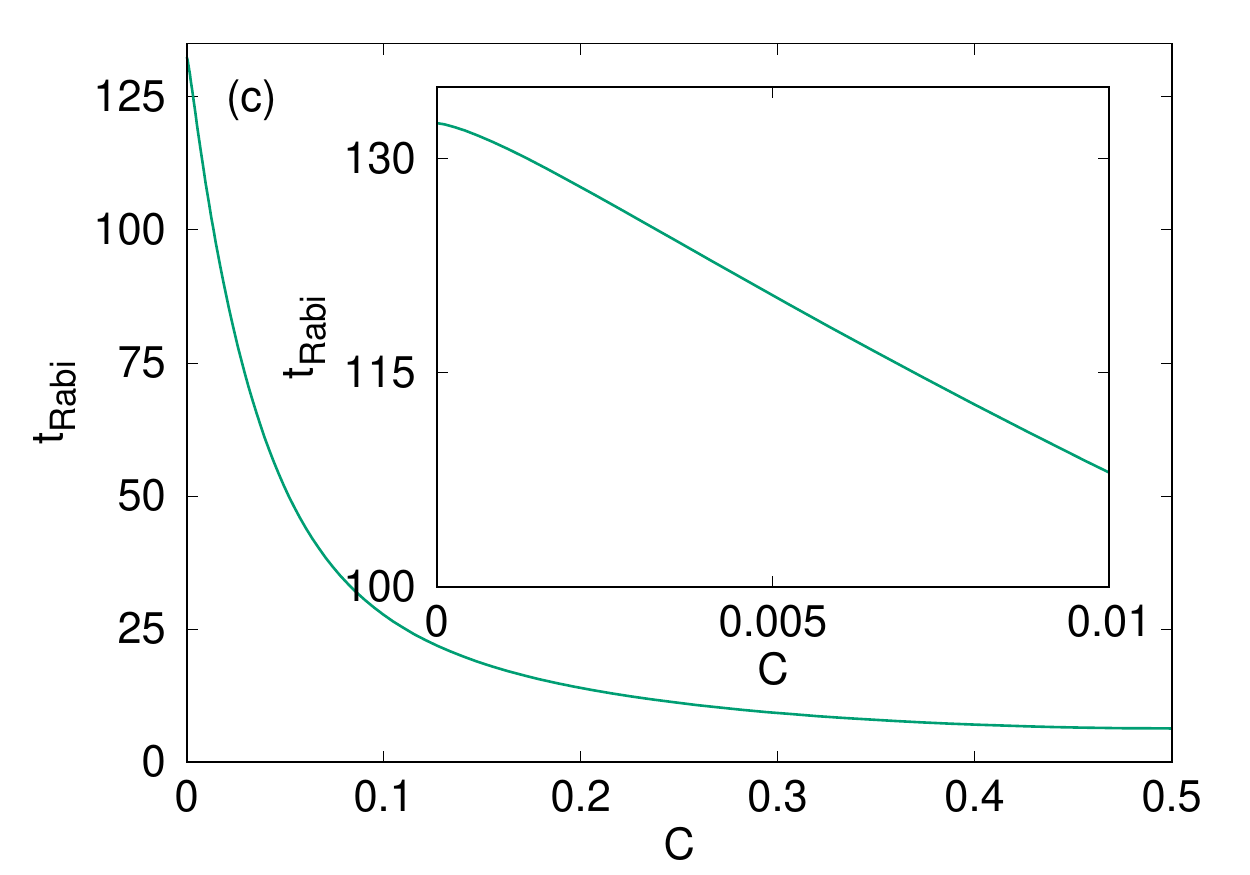}
\end{center} 
\caption{(a) An asymmetric double well potential with a small asymmetry $C=0.01$ (red solid curve) and its first six eigenfunctions. The symmetric double well (yellow dashed curve)
is also shown for comparison. A ten times magnified view of the relative positions of the energy levels with respect to the ground state is presented as the dotted horizontal lines while the horizontal solid curves represent a ten times magnified view of the corresponding eigenstates. 
The color code used for presenting the energy levels and the eigenstates is as follows: Magenta corresponds to the ground state, green to the first excited state, dark blue for the second excited state, dark yellow the third, sky blue the fourth and black presents the fifth excited state. (b) Ratio of the inter-band spacing to the intra-band spacing of the lowest band as a function of the asymmetry $C$ for the first and second higher band. The yellow curve represents $\frac{\Delta E_{52}}{\Delta E_{21}}$ while the sky blue curve represents the $\frac{\Delta E_{32}}{\Delta E_{21}}$. (c) $t_{Rabi}$ as a function of $C$. In the inset of panel (b) and (c), the range of of $C$ considered in this work is highlighted. See text for details. The quantities shown are dimensionless.}
\label{fig-trabi}
\end{figure}

Here we are interested in the dynamics of a system of $N$ interacting structureless bosons in a one-dimensional (1D) 
asymmetric double well which is governed by the time-dependent many-body Schr\"odinger equation:
\beq \label{MBSE}
\begin{split}
 \hat H \Psi = i \frac{\partial \Psi}{\partial t}, \qquad \hspace*{3cm}\\
 \hat H(x_1,x_2,\ldots,x_N) =  \sum_{j=1}^{N} \hat h(x_j) +  \sum_{k>j=1}^N W(x_j-x_k).
 \end{split}
 \eeq
Here $x_j$ {is} the coordinate of the $j$-th boson, $\hat h(x) = \hat T(x) + V_{T}(x)$ is the one-body Hamiltonian 
containing kinetic energy and trapping potential $V_{T}(x)$ terms, {and}
$W(x_j-x_k)$ is the pairwise interaction between the $j$-th and $k$-th bosons. 
Dimensionless units are employed throughout this work.
The asymmetric double well $V_{T}(x)$ 
is constructed by adding a linear slope of gradient $C$ to the symmetric double well which itself is obtained by fusing two slightly shifted harmonic potential 
$V_{L,R}=\frac{1}{2}(x\pm2)^2+Cx$, i.e.,
\begin{equation}
V_T(x) = \left\{
\begin{matrix}
\frac{1}{2}(x + 2)^2 + Cx, \hspace*{1cm} x < -\frac{1}{2} \cr 
%\vspace*{0.3cm}
\frac{3}{2}(1-x^2) + Cx, \hspace*{1cm} |x| \le \frac{1}{2} \cr 
%\vspace*{0.3cm}
\frac{1}{2}(x - 2)^2 + Cx, \hspace*{1cm} x > \frac{1}{2} \cr  
\end{matrix}
\right.\,.
\end{equation}
{The symmetric double-well part is taken from ~\cite{Sudip2018}. This will allow us to relate and compare results in the asymmetric junctions to that in the symmetric one. The shape of an asymmetric} {double well for $C=0.01$ used in this work along with its first few energy levels $E_n$ and eigenstates $\varphi_n(x)$ are shown in Fig.~\ref{fig-trabi}(a). One can see that it is hardly distinguishable from the symmetric double well. Also, even for such a small asymmetry $C$, the superposition of the first two eigenstates are not completely localized in one or the other well, thereby affecting the density oscillations between the two wells. Moreover, the spacing between the two successive energy levels increases as one goes up the spectrum: while the lowest two energy levels lie very close to each other and form the lowest energy band, the higher energy levels from $E_5$ onward are practically unaffected by the barrier between the two wells and form an almost uniform spectrum. Therefore, the dynamics of the system in such an asymmetric double well is primarily controlled by the lowest energy band. However, with increasing $C$, the spectrum starts to be affected more prominently by the asymmetry and the higher energy levels begin to play more important role in the dynamics. }

{To highlight the point further, we compute the ratio of the inter-band spacing to the intra-band spacing of the lowest band, viz., $\frac{\Delta E_{n2}}{\Delta E_{21}}=\frac{E_n-E_2}{E_2-E_1}$. In Fig.~\ref{fig-trabi}(b), we explicitly show the ratios $\frac{\Delta E_{32}}{\Delta E_{21}}$ and $\frac{\Delta E_{52}}{\Delta E_{21}}$ as functions of $C$. We see that starting from a relatively large value for the symmetric double well ($C=0$), these ratios decay rapidly with $C$ with the decay rate being higher for $\frac{\Delta E_{52}}{\Delta E_{21}}$. This implies that coupling to higher energy levels starts to grow with increasing $C$. For the range of values of $C$ of our interest [shown in the inset of Fig.~\ref{fig-trabi}(b)], we see that the $\frac{\Delta E_{32}}{\Delta E_{21}}$ and $\frac{\Delta E_{52}}{\Delta E_{21}}$ are quite large and of the order of their values for the $C=0$. Therefore, in the regime of our interest, the lowest energy band is expected to play the lead role in the dynamics of the system. However, the next nearest band may influence the dynamics by giving rise to the breathing mode oscillations on top of the Rabi oscillations controlled primarily by the lowest band. Also,
in our present study, the inter-atomic interaction $W(x_j-x_k)$ may lead to a coupling with the higher energy levels. 
Therefore, even for such a small asymmetry, it is necessary to effectively take all bands into account. Only then, one can be sure that the lowest band is the dominant one.}

%Figure 1

{The time period of the Rabi oscillations in the double well, $t_{Rabi}=\frac{2\pi}{E_2-E_1}$, provides a natural choice for the time scale of the dynamics.  $t_{Rabi}$ as a function of the asymmetry $C$ is shown in Fig.~\ref{fig-trabi}(c) with the region of our interest being highlighted in the inset. We note that $t_{Rabi}$ also decreases exponentially with $C$. Actually, for a small asymmetry $C$, the ground state and the first excited state in the asymmetric double well are delocalized, and
$t_{Rabi}$ gives the time period of Rabi oscillations in the double well. However,
for large $C$, the barrier becomes very high and there is no tunneling back and forth between the two wells. Then $E_1$ and $E_2$ become 
the lowest two energy levels in the lower well $V_L(x)$ and $\frac{2\pi}{E_2-E_1}$ 
{is associated with the} time period of breathing mode oscillations. 
In this work, as already mentioned above, we consider only small asymmetries and therefore, we will use the time period of Rabi oscillations $t_{Rabi}$ as a unit of time for the 
description of the dynamics in a particular asymmetric double well trap. However, as shown above, $t_{Rabi}$ varies with $C$ and therefore is not suitable for comparing the dynamics
in different asymmetric traps. However, from the inset of Fig~\ref{fig-trabi}(c), we note that for the range of values of $C$ of our interest $t_{Rabi} \sim 10^{-2}$ and, therefore, for comparing the dynamics in different traps, we will use $t_0=100$ as a unit of time.}

Further, it is convenient to {define the different quantities of interest} in terms of the one-body and the 
two-body reduced density matrices~\cite{Lowdin,Yukalov,Mazz,RDMs} 
instead of the full many-body wavefunction. 
Given the normalized many-body wavefunction $\Psi(t)$, the reduced one-body density matrix can be calculated as
%Eqn(6)
\beqn\label{1RDM}
\rho^{(1)}(x_1|x_1^{\prime};t) & = &
N \int d x_2 \ldots d x_N \, \Psi^\ast(x_1^{\prime},x_2,\ldots,x_N;t)  \nonumber \\
& \, & \times \Psi(x_1,x_2,\ldots,x_N;t) \nonumber \\
& = &\sum_{j=1}^{M} n_j(t) \, \phi^{\ast{NO}}_j(x_1^{\prime},t)\phi^{NO}_j(x_1,t).
\eeqn
Here, $\phi^{NO}_j(x_1,t)$ are the time-dependent natural orbitals and $n_j(t)$ the time-dependent natural occupation numbers. 
The natural occupations $n_j(t)$ are used to characterize the {(time varying)} degree of condensation 
in a system of interacting bosons \cite{PeO56} and satisfy $\sum_{j=1}^{M} n_j = N$ ($M$ is the number of single particle orbitals used to construct the many-boson wavefunction, 
see Sec.~\ref{MCTDHB}). 
If only one {macroscopic} eigenvalue 
$n_1(t) \approx {\mathcal O}(N)$ exists, the system is condensed \cite{PeO56} whereas if 
there are more than one {macroscopic} eigenvalues, the BEC
is said to be fragmented {\cite{MCHB,NoS82,No96,Spekkens99,Ueda,RDMs}}. The diagonal of the $\rho^{(1)}(x_1|x_1^{\prime};t)$ gives the density of the system
$\rho(x;t) \equiv \rho^{(1)}(x|x^{\prime}=x;t) \label{1RDMdiag}$.

Similarly, the two-body density can be calculated as  
\begin{equation}\label{2RDM}
\begin{split}
\rho^{(2)}(x_{1},x_{2} \vert x_{1}^{\prime}, x_{2}^{\prime};t)= \hspace*{4cm}\\
N(N-1)\int_{}^{}d x_{3} \ldots d x_{N} \Psi^{*}(x_{1}^{\prime},x_{2}^{\prime},x_{3},\ldots,x_{N};t) \\
\times \Psi(x_{1},x_{2},x_{3}, \ldots, x_{N};t).
\end{split}
\end{equation} 
Therefore, the matrix elements of the two-body reduced density matrix are given by 
$\rho_{ksql}=\left<\Psi\left|b_k^\dag b_s^\dag b_q b_l\right|\Psi\right>$ where $b_k$ and $b_k^\dag$
are the bosonic annihilation and creation operators, respectively.

\subsection{Physical quantities}
In this work, we will study the dynamics of the system by exploring the time evolution of different physical quantities defined as follows. 
While some of these quantities can be studied both at the mean-field and the many-body levels, others can only be studied at the many-body level.  

\begin{enumerate}
 \item[a)] \textit{Survival probability}. In the dynamics of BEC in an asymmetric double well following a trapping quench from a harmonic well 
to an asymmetric double well at $t=0$, we can prepare the initial BEC state either in the left well (L) or in the right well (R). 
Accordingly, we can calculate two {types} of survival probabilities. For starting with the initial BEC state in the left well, 
we can define the survival probability in the 
left well [$p_L(t)$] as 
\begin{equation}
 p_L(t)=\int_{-\infty}^0 {\rm d}x \frac{\rho_L(x;t)}{N},
\end{equation}
where $\rho_L(x;t)$ is the density {when the initial BEC state is prepared} in the left well. Similarly, when the initial state is prepared in the right well, 
the survival probability in the right well [$p_R(t)$] can be 
defined as 
\begin{equation}
 p_R(t)=\int_0^{\infty} {\rm d}x \frac{\rho_R(x;t)}{N},
\end{equation}
where $\rho_R(x;t)$ is the density {when the initial condensate is} in the right well. For a symmetric double well, both  $\rho_R(t)$ and $\rho_L(t)$ are 
equivalent and therefore we have only {a single type of} survival probability $p(t)$, i.e., $p_L(t) \equiv p_R(t) = p(t)$. {Since the density can be studied both at the mean-field and the many-body levels, survival probabilities can also be calculated both at the mean-field and the many-body levels.}
 
\item[b)] \textit{Depletion and fragmentation}. As discussed above, when $n_1(t) \approx {\mathcal O}(N)$ the system is condensed and the sum over all the microscopic fractions of
occupations in the higher orbitals {$f=\sum_{j=2}^{M} \frac{n_j}{N}$} ($M$ is the number of orbital, see above) 
is known as the depletion {per particle}. On the other hand for a fragmented system, the macroscopic
{occupation} of a higher natural orbital, 
viz. $f=\frac{n_{j>1}}{N}$ where $n_j \approx {\mathcal O}(N)$, is called fragmentation. {From the definition, it is clear that one needs more than one orbital to study
the depletion and fragmentation and hence, these quantities can only be calculated by at least a two-orbital many-body theory and preferably a multi-orbital many-body theory. 
We remark that the depletion of a BEC is usually small and may not have a prominent effect on the density per particle and energy per particle which, in effect, can be 
accurately described by a mean-field theory. However, fragmentation can have a dominant effect on the energy per particle and the density per particle of the system. Moreover,
though the depletion and the fragmentation are physically different quantities and appear under different conditions, for a two-orbital theory they have the same mathematical
expression. Accordingly, for the computation with $M=2$ orbitals only we will refer to both of them by $f$, see Sec.~\ref{MCTDHB} below.}

\item[c)] \textit{many-particle position and momentum variance.} The quantum variance of an observable is a fundamental quantity in quantum mechanics due to its 
connection with the uncertainty principle.
It gives a measure of the quantum resolution with which an observable can be measured. %It can also be measured experimentally in ultra-cold atomic systems~\cite{}. 
For any many-body operator $\hat A=\sum_{j=1}^N \hat a(x_j)$ where $\hat a(x_j)$ is a Hermitian operator and local in position space, the variance per particle 
$\frac{1}{N}\Delta_{\hat A}^2(t)$~\cite{Klaiman2016,Klaiman2015,Marcus,Klaiman2018} is given by
\begin{widetext}
\beqn\label{dis}
& & \frac{1}{N}\Delta_{\hat A}^2(t)  = \frac{1}{N} 
\left[\langle\Psi(t)|\hat A^2|\Psi(t)\rangle - \langle\Psi(t)|\hat A|\Psi(t)\rangle^2\right]
 \equiv \Delta_{\hat a, density}^2(t) + \Delta_{\hat a, MB}^2(t), \nonumber \\
& & \quad \quad \Delta_{\hat a, density}^2(t) = 
\int d x \frac{\rho(x;t)}{N} a^2(x) - \left[\int dx \frac{\rho(x;t)}{N} a(x) \right]^2, \nonumber \\ 
& & \quad \quad \Delta_{\hat a, MB}^2(t) = \frac{\rho_{1111}(t)}{N} \left[\int d x |\phi^{NO}_1(x;t)|^2 a(x) \right]^2
 - (N-1) \left[\int dx \frac{\rho(x;t)}{N} a(x) \right]^2 + \nonumber \\
& &  \sum_{jpkq\ne 1111}\frac{\rho_{jpkq}(t)}{N} \left[\int d x \phi^{\ast{NO}}_j(x;t) \phi^{NO}_k(x;t) a(x) \right] 
\left[\int dx \phi^{\ast{NO}}_p(x;t) \phi^{NO}_q(x;t) a(x)\right]. \
\eeqn
\end{widetext}
Here the first term, $\Delta_{\hat a, density}^2(t)$, is the variance of $\hat a(x)$ resulting from the density per particle $\frac{\rho(x;t)}{N}$, whereas the second term, 
$\Delta_{\hat a, MB}^2(t)$,
takes into account all other contributions to the many-particle variance. {Similar expressions hold for operators 
which are local in momentum space. We point out that one can, in principle, study 
the variance of any operator at the mean-field level 
by substituting the many-body wavefunction $\Psi(t)$ by the corresponding mean-field wavefunction. % in Eq.~(\ref{dis}). 
However, in the mean-field theory, only $\Delta_{\hat a, density}^2(t)$ has a nonzero contribution while
$\Delta_{\hat a, MB}^2(t)$ is identically equal to zero. Therefore, even for the interaction strengths for which the mean-field theory is expected to accurately describe the 
density per particle of the system, the many-body variance can deviate from its mean-field result. Accordingly, in this work we will consider the variances
of the many-particle position and momentum operators at the many-body level only.} 
\end{enumerate}

\subsection{{Computational Method}}
\label{MCTDHB}
The time-dependent many-boson Schr\"odinger equation (\ref{MBSE}) cannot be solved exactly (analytically), 
except for a few specific cases only, see, e.g.,~\cite{Marvin}.
Hence, {to solve Eq.~(\ref{MBSE}) in-principle numerically exactly, the multi-configurational time-dependent Hartree method for bosons
(MCTDHB)}{,}~\cite{Streltsov2007,Ofir2008} was developed and 
{benchmarked with an exactly-solvable model~\cite{Lode2012,Axel_MCTDHF_HIM}. This method has already} been extensively used in the literature 
\cite{Sakmann2009,MCTDHB_OCT,MCTDHB_Shapiro,Tunneling_Rapha,Kota2015,Axel2016,PRA2016Axel,Axel2017,Cosme2017,Sudip2018,Rohmbik}.
Detailed derivation of the MCTDHB equation of motions can be found in~\cite{Ofir2008}. Below we briefly describe the basic idea behind the method.

In MCTDHB, the ansatz for solving Eq.~(\ref{MBSE}) {is obtained by the superposition of all possible
$\begin{pmatrix} 
N+M-1\\
N
\end{pmatrix}$
configurations, obtained by distributing $N$ bosons in $M$ time-dependent single{-}particle states $\phi_k(x,t)$, which we call orbitals, i.e,}
\beq\label{MCTDHB_Psi}
\left|\Psi(t)\right> = 
\sum_{\vec{n}}C_{\vec{n}}(t)\left|\vec{n};t\right>,
\eeq
where the occupations $\vec{n}=(n_1,n_2,\cdots,n_M)$ preserve the total number of bosons $N$. {For an exact theory, $M$ should be infinitely large. However, for numerical 
computations one has to truncate the series at a finite $M$. In actual calculations, we keep on increasing $M$ until we reach the convergence with respect to $M$ and thereby we obtain
a numerically-exact result. In the context of bosons in a double-well, the latter implies that the MCTDHB theory effectively takes all required bands into account. 
Here we would like to point out that for 
$M=1$, the ansatz Eq.~(\ref{MCTDHB_Psi}) gives back the ansatz for the Gross Pitaevskii theory~\cite{rev1}.} 

{Therefore, solving for the time-dependent wavefunction $\Psi(t)$ boils down to  
the determination of the time-dependent coefficients $\{C_{\vec{n}}(t)\}$ 
and the time-dependent orbitals $\{\phi_k(x,t)\}$.}
Employing the usual Lagrangian formulation of the time-dependent 
variational principle \cite{LF1,LF2} subject to the orthonormality 
between the orbitals, 
{the working equations of the MCTDHB are obtained as follows} 
\beqn\label{MCTDHB1_equ}
i\left|\dot\phi_j\right> & = & \hat {\mathbf P} \left[\hat h \left|\phi_j\right>  + \sum^M_{k,s,q,l=1} 
  \left\{\brho(t)\right\}^{-1}_{jk} \rho_{ksql} \hat{W}_{sl} \left|\phi_q\right> \right]; \nonumber\\
%& &  
\qquad \hat {\mathbf P} &=& 1-\sum_{j^{\prime}=1}^{M}\left|\phi_{j^{\prime}}\left>\right<\phi_{j^{\prime}}\right| \nonumber\\
& & {\mathbf H}(t)\C(t) = i\frac{\partial \C(t)}{\partial t}.
\eeqn
Here, $\brho(t)$ is the reduced one-body density matrix [Eq.~(\ref{1RDM})], 
$\rho_{ksql}$ are the elements of the two-body reduced density matrix [Eq.~(\ref{2RDM})], and ${\mathbf H}(t)$ is the Hamiltonian matrix {with the elements}
$H_{\vec{n}\vec{n}'}(t) = \left<\vec{n};t\left|\hat H\right|\vec{n}';t\right>$.
A parallel version of MCTDHB has been
implemented using a novel mapping technique~\cite{Streltsov1, Streltsov2}. We {mention} that by propagating in
imaginary time the MCTDHB equations also allow {one} to determine {the ground %and excited
state} of interacting many-boson systems, {see~\cite{MCHB,Lode2012}}. In our present work we have performed all computations with $M=2$ time-adaptive orbitals. 
{By repeating our computations with $M=4,6$, and $8$ orbitals the results have been verified and 
found to be highly accurate for the quantities and propagation {times} considered here.
Further details about our numerical computations and its convergence are discussed in {the} 
Appendix.}

\section{Results}
\label{results}

In this section, we discuss the outcome of our investigation of the dynamics of a BEC in an asymmetric double well. Specifically, we are interested to understand how the presence of 
asymmetry influences the overall dynamics of the BEC for different interaction {strengths}. 
In this work, we consider the dynamics of systems made of $N=100-10000$ bosons interacting via a contact $\delta$ interaction of strength $\lambda_0$ which 
corresponds to the interaction parameter $\Lambda = \lambda_0(N-1)$.
We {again} remind that for an asymmetric double well trap, one can prepare the initial state either in the left well $V_L(x)$ or in the right well $V_R(x)$, and 
then allow the system to evolve in time in the double well $V_{T}(x)$. Accordingly, we will study the {dynamics of the system}
once starting from $V_L(x)$ and then from $V_R(x)$.

\subsection{Quantum dynamics in an asymmetric double-well}
\label{dynamics}
As already mentioned above, we will characterize the dynamics in an asymmetric double well trap 
by the time evolution of a few physical quantities such as the survival probability, depletion and fragmentation, and the many-particle position and momentum variances. 
The corresponding dynamics in the symmetric double well will serve as a reference for our {analysis}. {We studied the time evolution of these quantities at the many-body levels for a weak as well as a strong interaction strength $\Lambda$. We also studied the corresponding dynamics at the mean-field level, wherever applicable, to explicitly highlight the many-body effects in the dynamics.}

\subsubsection{Survival probability}
\label{pl}

\begin{figure}[!ht]
\begin{center}
\begin{tabular}{cc} 
\includegraphics[width=0.5\linewidth]{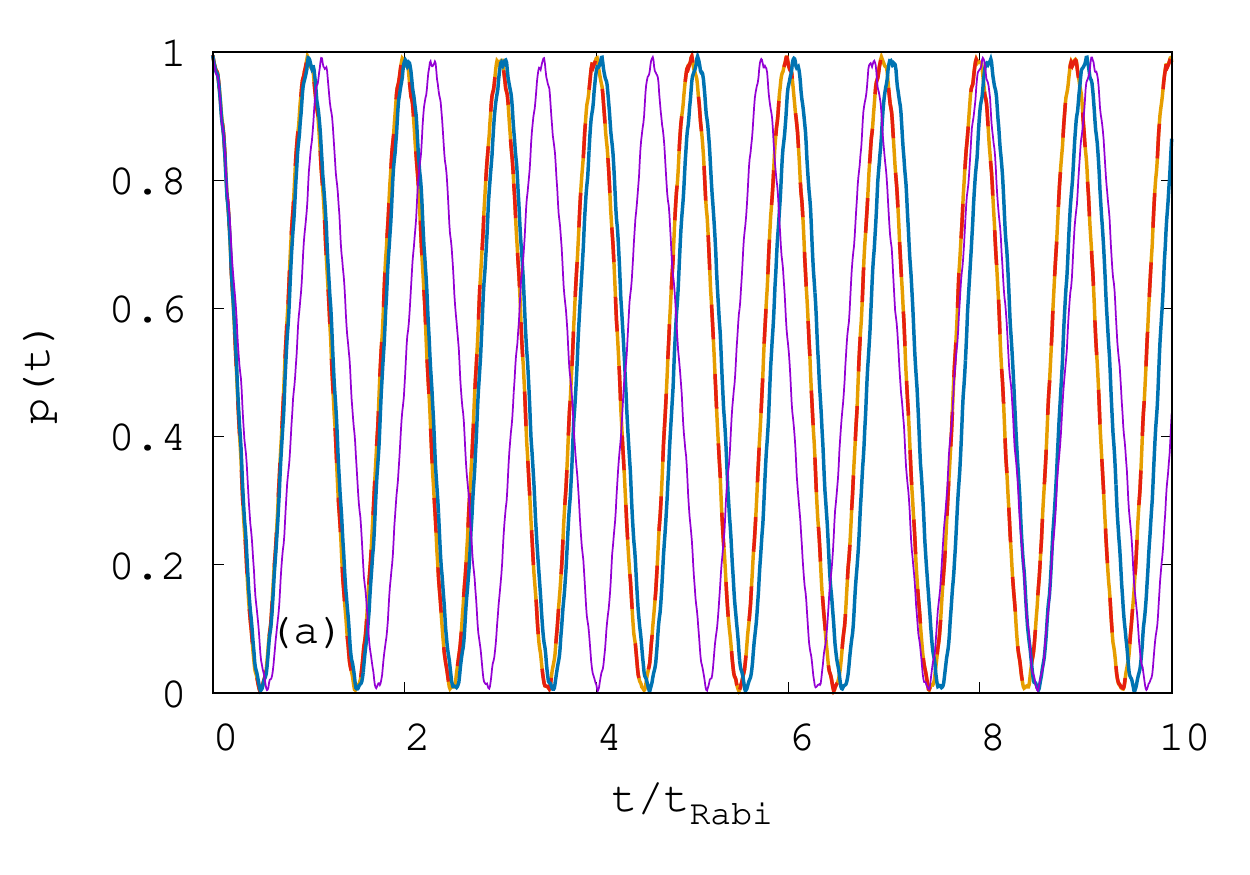} & 
\includegraphics[width=0.5\linewidth]{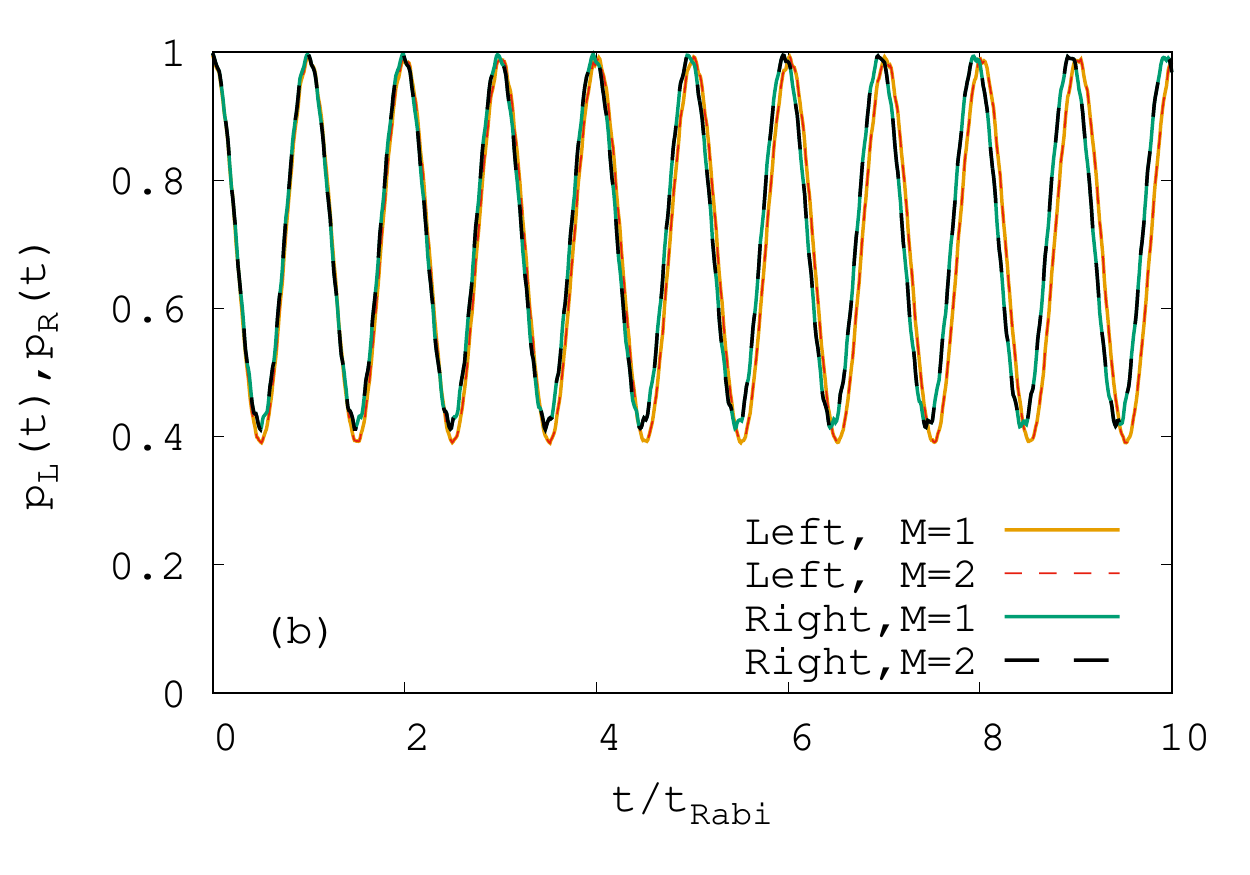} \\
\includegraphics[width=0.5\linewidth]{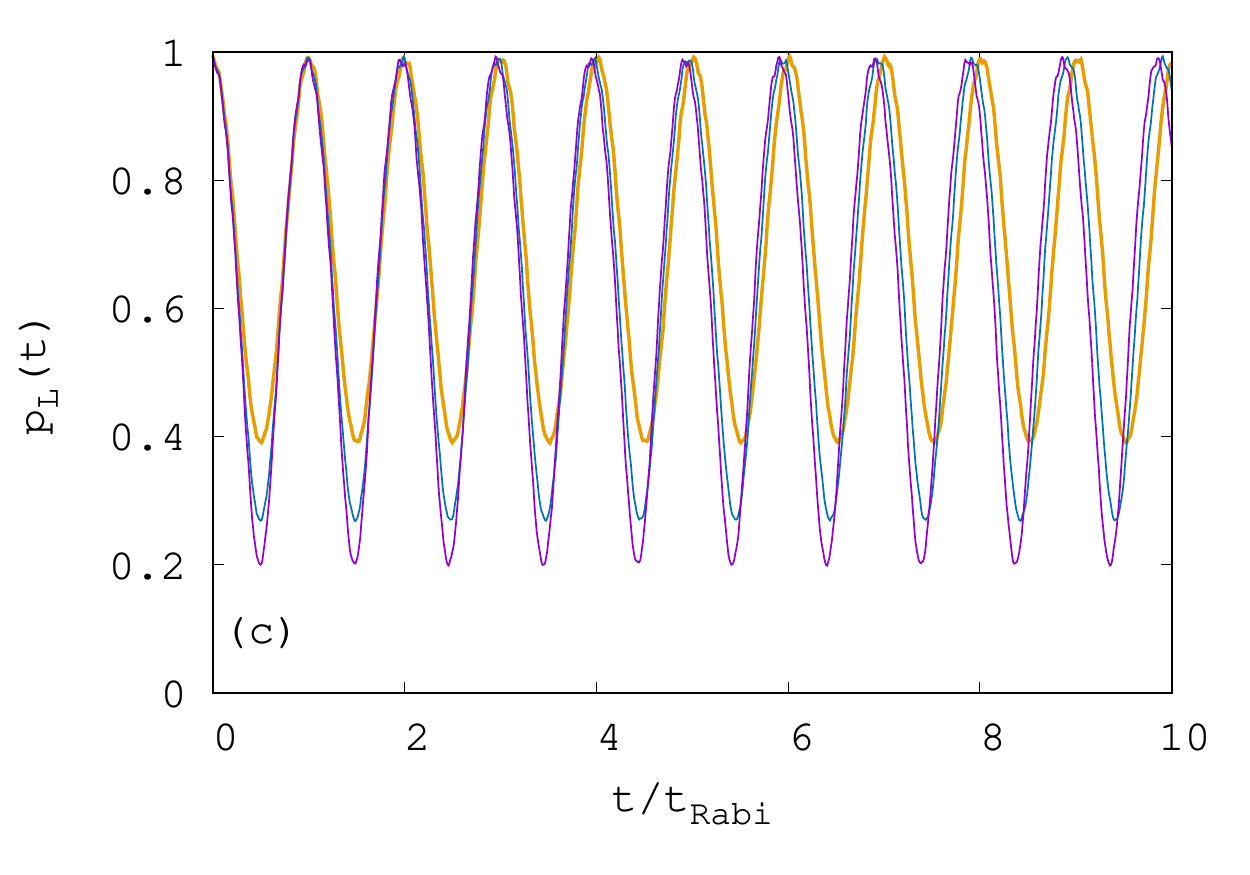} &
\includegraphics[width=0.5\linewidth]{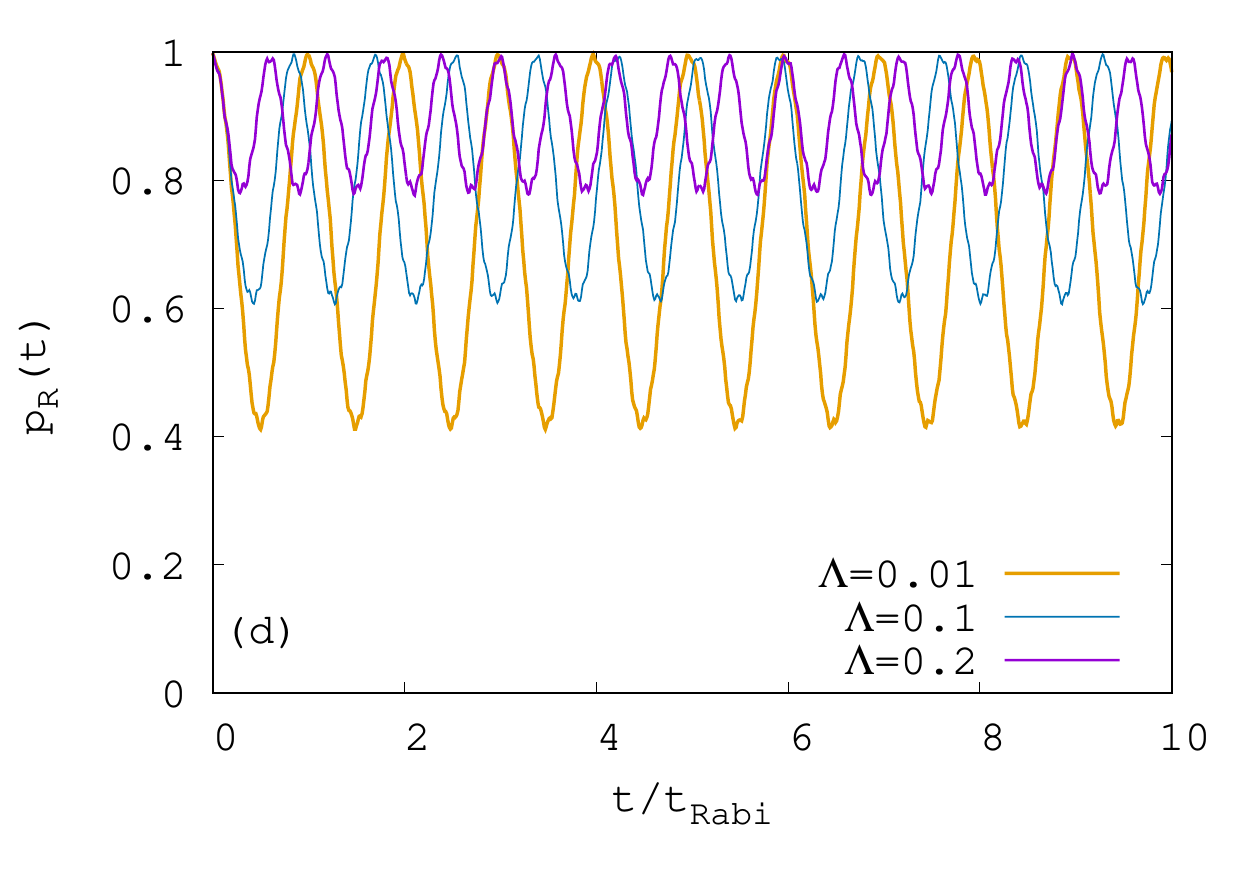}  \\
\end{tabular}
\end{center}
\caption{(a) Time evolution of the survival probability $p(t)$ in the left well of a symmetric double well for different interaction strengths $\Lambda$. 
Mean-field results of $p(t)$ for $\Lambda=0.01$ (largest amplitude), $\Lambda=0.1$ (intermediate amplitude), and $\Lambda=0.2$ (smallest amplitude) 
correspond to the yellow, blue, and magenta smooth curves, respectively. 
The MCTDHB result of $p(t)$ computed with $M=2$ orbitals for a system of $N=1000$ bosons and $\Lambda=0.01$ is shown as the red dashed curve.  
(b) Time evolution of the survival probabilities in the left [$p_L(t)$] and right [$p_R(t)$] well of an asymmetric double well with asymmetry $C=0.01$ for $\Lambda=0.01$. 
The yellow smooth curve corresponds to the mean-field result of $p_L(t)$ while the red dashed curve represents the MCTDHB result of $p_L(t)$ computed with $M=2$ orbitals 
for a system of $N=1000$ bosons. On the other hand, the green smooth curve represents the corresponding mean-field result of $p_R(t)$ while the corresponding MCTDHB result 
computed with $M=2$ orbitals for a system of $N=1000$ bosons is shown as the black dashed curve. (c) Mean-field results of the time evolution of the survival probability in 
the left [$p_L(t)$] well of an asymmetric double well with asymmetry $C=0.01$ for different $\Lambda$. Color codes are explained in panel (d).
(d) The corresponding mean-field time evolution of the the survival probability in the right well [$p_R(t)$]. The quantities shown here are dimensionless.}
\label{fig-pl}
\end{figure}

{We start with the survival probability $p(t)$ in the initial well (left well) of a symmetric double well which will serve as the reference for our subsequent analysis 
of the survival probabilities $p_L(t)$ and $p_R(t)$ in an asymmetric double well. As discussed above, the survival probabilities can be studied both at the  mean-field and 
many-body levels. Accordingly, in Fig.~\ref{fig-pl}(a) we plot the mean-field results of $p(t)$ for different $\Lambda$. We see that $p(t)$ performs smooth oscillations back and forth
between the two wells. %Such oscillations are fundamental to the dynamics in a symmetric double well. 
For a symmetric double well, the one-body Hamiltonian ${\hat h}(x)$ is 
invariant under parity and therefore its eigenstates are also parity eigenstates: the ground state has even parity while the first excited state is odd. Accordingly, 
the superpositions of these two states are localized in one or the other well. Therefore, when a one-particle state initially localized in one well is allowed to evolve in time, 
it keeps on tunneling back and forth between the two wells. However, in case of systems with a finite number of interacting particles like a BEC, there will be an effect of 
inter-particle interactions on this tunneling dynamics. Such effects are manifested through the frequency of oscillations of $p(t)$ in Fig.~\ref{fig-pl}(a). We observe that, as
the inter-atomic interaction $\Lambda$ increases, the frequency of oscillations of $p(t)$ decreases. In the same figure, we also plot the many-body results of $p(t)$ for $N=1000$,
$\Lambda=0.01$, and $M=2$ orbitals.}
The complete overlap between the mean-field and the many-body results of $p(t)$ confirms that 
{for these parameters, the density per particle of the system and hence the survival probability can be accurately described by the mean-field theory}.

{Next, we consider an asymmetric double well potential with a very small asymmetry, $C=0.001$. Due to the presence of asymmetry, the parity symmetry of ${\hat h}(x)$ 
is now lifted and therefore, the eigenstates of ${\hat h}(x)$ are no more parity eigenstates. Accordingly, the superpositions of the first two eigenstates of ${\hat h}(x)$ (see 
Fig.~1) are no longer well localized in one or the other well. Therefore, if a one-particle state initially localized in one well is allowed to evolve in time, 
it will become partially delocalized 
over both wells and hence there will never be full oscillations of the density of the system between the two wells.}

{However, for such a small asymmetry $C=0.001$ and a weak interaction $\Lambda=0.01$, we did not find any visible suppression of oscillations of $p_L(t)$ and $p_R(t)$
at the mean-field level (not shown here). 
Moreover, both $p_L(t)$ and $p_R(t)$ would lie on top of each other.  Therefore, we conclude that $C=0.001$ is too small of an asymmetry to have any visible impact on the
tunneling dynamics of the system. We also 
repeat our calculations of $p_L(t)$ and $p_R(t)$ for a system of $N=1000$ interacting bosons by the MCTDHB method with $M=2$ orbitals and 
confirm that these mean-field descriptions of $p_L(t)$ and $p_R(t)$ are accurate.}

%Figure 2

Next, we enhance the asymmetry to $C=0.01$ keeping $\Lambda=0.01$ fixed. The mean-field $p_L(t)$ and $p_R(t)$ are shown in Fig.~\ref{fig-pl}(b). 
{Now, we observe the expected suppression of tunneling between the two wells. The amplitudes of oscillations of both $p_L$ and $p_R$ have decreased 
by nearly $40\%$ indicating that almost $40\%$ of the system does not tunnel out of the initial well.} Moreover, though $p_L(t)$ and $p_R(t)$ practically overlap with each other, 
a small phase difference is found to develop with time after a few oscillations. {This small phase difference is 
the combined effect of asymmetry and such weak interaction on the dynamics.} 
In the same figure, we also plot the many-body $p_L(t)$ and $p_R(t)$ obtained with $M=2$ orbitals. That the respective mean-field and many-body curves for 
$p_L(t)$ and $p_R(t)$ again lie atop each other confirms that the {mean-field description of the system is accurate for such a weak $\Lambda$.
Thus, here we observe that even an asymmetry as small as $C=0.01$ has a prominent effect on the macroscopic tunneling dynamics of the system.}

To {further probe the effect of interaction on the tunneling dynamics between the two wells of an asymmetric double well,} 
we next increase $\Lambda$, keeping the asymmetry $C=0.01$ fixed. 
The mean-field results of $p_L(t)$ and $p_R(t)$ for different $\Lambda$ are shown in Fig.~\ref{fig-pl}(c) and (d), respectively. We find a complementary effect of $\Lambda$ on 
$p_L(t)$ and $p_R(t)$ at the mean-field level. While both $p_L(t)$ and $p_R(t)$ still oscillates back and forth, their amplitudes and frequencies vary with $\Lambda$ in 
an opposite fashion. {While stronger $\Lambda$ facilitates oscillations of $p_L(t)$, it suppresses the oscillations of $p_R(t)$. Also, the frequencies of oscillations
are found to decrease with increasing $\Lambda$ for starting the dynamics from the left well, whereas when started from the right well, 
the frequencies increase with increasing $\Lambda$.}

{Qualitatively, the repulsive interaction pushes up the energy of the BEC with respect to the barrier height. 
For a sufficiently high barrier, the energy levels of the 
ground state and the first excited state of an asymmetric double well approximately coincide with the ground states of the left (lower) and the right (upper) wells, respectively. 
Therefore, when the initial state is prepared in the right (upper) well, the energy of the initial condensate becomes closer to the energy level of the 
first excited state of the asymmetric double well with increasing repulsive interaction $\Lambda$. 
Thus, the system tends more to remain in that state and the tunneling is more and more suppressed with increasing $\Lambda$.
On the other hand, when the initial condensate is prepared in the left (lower) well, its energy is pushed away from the ground state of the asymmetric double well by the 
repulsive interaction and therefore, it becomes more prone to tunneling with increasing $\Lambda$}

\begin{figure}[!ht]
\begin{center}
\begin{tabular}{cc} 
\includegraphics[width=0.50\linewidth]{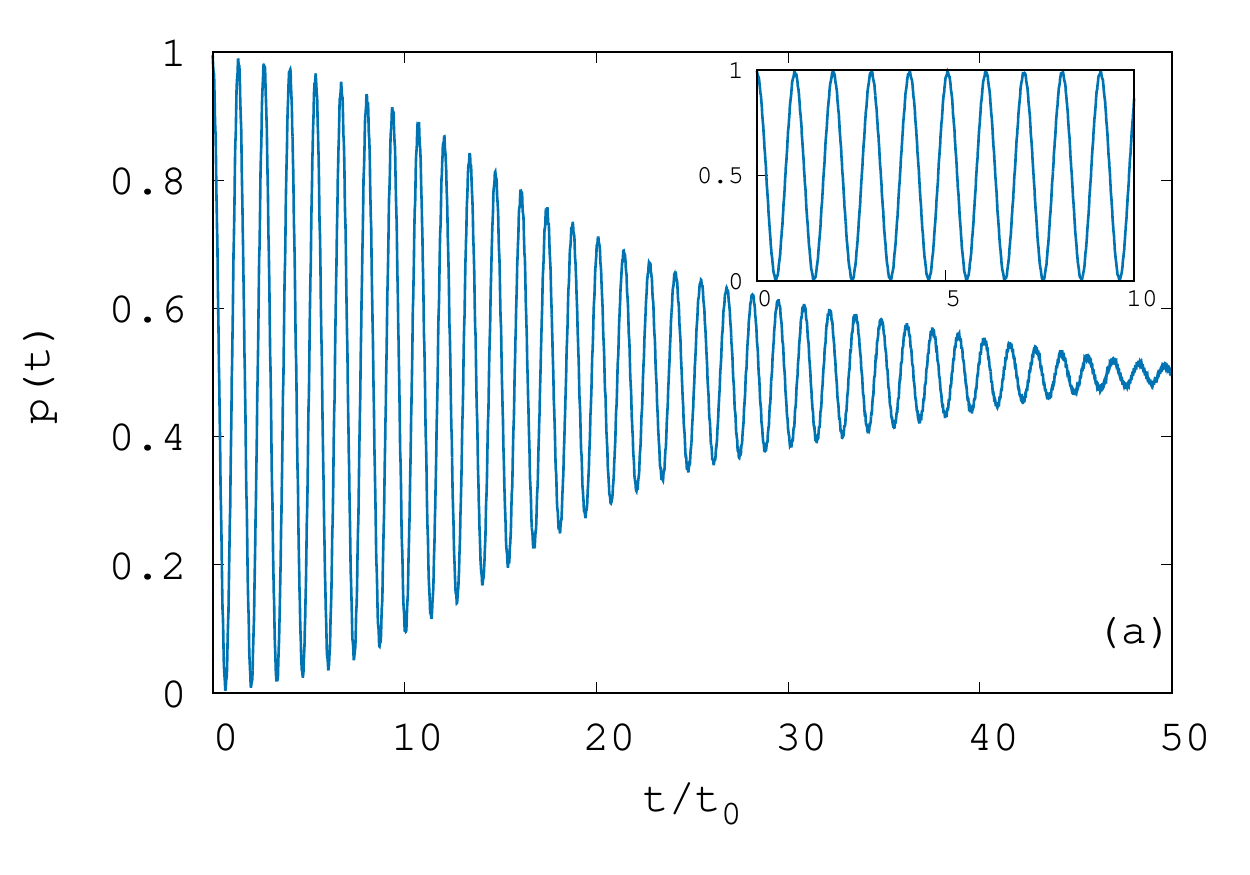} & 
\includegraphics[width=0.50\linewidth]{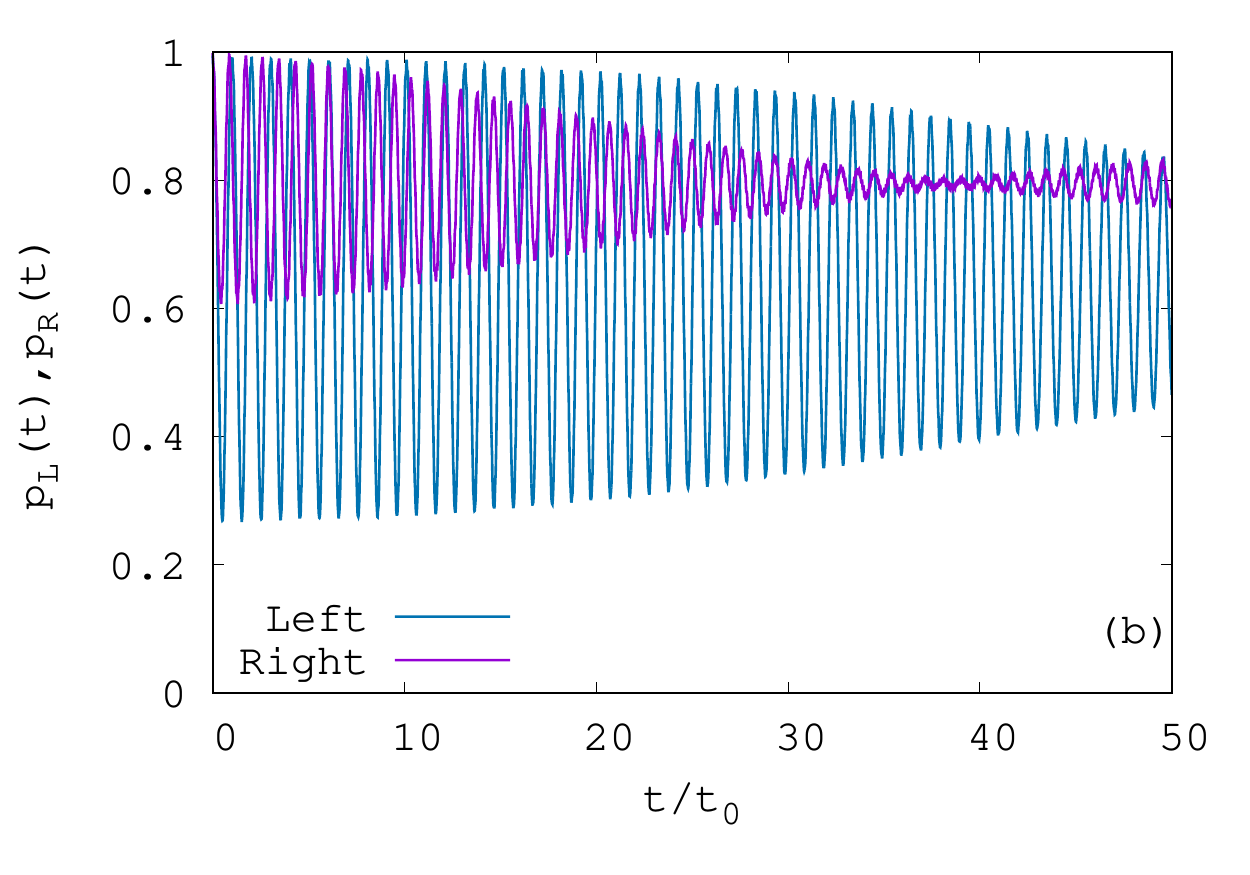}   \\
\end{tabular}
\end{center}
\caption{(a) MCTDHB result of the time evolution of the survival probability $p(t)$ in the left well of a symmetric double well computed with $M=2$ orbitals for a system of 
$N=1000$ bosons and $\Lambda=0.1$. The corresponding mean-field result is shown in the inset for comparison. (b) Corresponding MCTDHB result of the time evolution of the survival 
probabilities in the left [$p_L(t)$] and right [$p_R(t)$] well of an asymmetric double well with asymmetry $C=0.01$. We used $M=2$ orbitals for the computations of both 
$p_L(t)$ and $p_R(t)$ which are shown as the blue and magenta curves, respectively, as explained in the figure itself. The quantities shown are dimensionless.}
\label{fig-pl-lp1}
\end{figure}

At stronger $\Lambda$ keeping $N$ fixed, a mean-field theory may not be sufficient to describe the system.
Accordingly, we again refer to the symmetric double well case. Fig.~\ref{fig-pl-lp1}(a) exhibits the time evolution of $p(t)$ for a symmetric double well calculated by 
MCTDHB with $M=2$ orbitals. In the inset, the corresponding mean-field 
result of $p(t)$ is provided for comparison. We clearly see that, contrary to the mean-field result, the many-body result for $p(t)$ exhibits a collapse of the oscillations, 
thereby making a many-body calculation necessary for $\Lambda \ge 0.1$ for a system of $N=1000$ bosons. Therefore, next, we calculate the $p_L(t)$ and 
$p_R(t)$ by MCTDHB method with $M=2$ orbitals for the same parameters as in Fig.~\ref{fig-pl}(c) and (d) for $N=1000$ bosons.  
As an example, here we present the many-body results only for $\Lambda=0.1$ in Fig.~\ref{fig-pl-lp1}(b). 
The collapse of oscillations for both $p_L(t)$ and $p_R(t)$ can be seen on top of the overall mean-field 
effects described above. However, the collapse time differs: While the collapse for $p_L(t)$ is quicker compared to the symmetric double well, it is delayed for $p_R(t)$. 
We further found that, with an increase in $\Lambda$, the 
collapse is quicker for both the $p_L(t)$ and $p_R(t)$ for a fixed asymmetry $C$. On the other hand, for a fixed $\Lambda$, the collapse of both $p_L(t)$ and $p_R(t)$ is deferred 
with increasing $C$ in terms of the number of {Rabi cycles (recall that $t_{Rabi}$ depends on $C$).}

%Figure 3

\subsubsection{Depletion and fragmentation}
\label{fragmentation}

\begin{figure}[!ht]
\begin{center}
\includegraphics[width=0.90\linewidth]{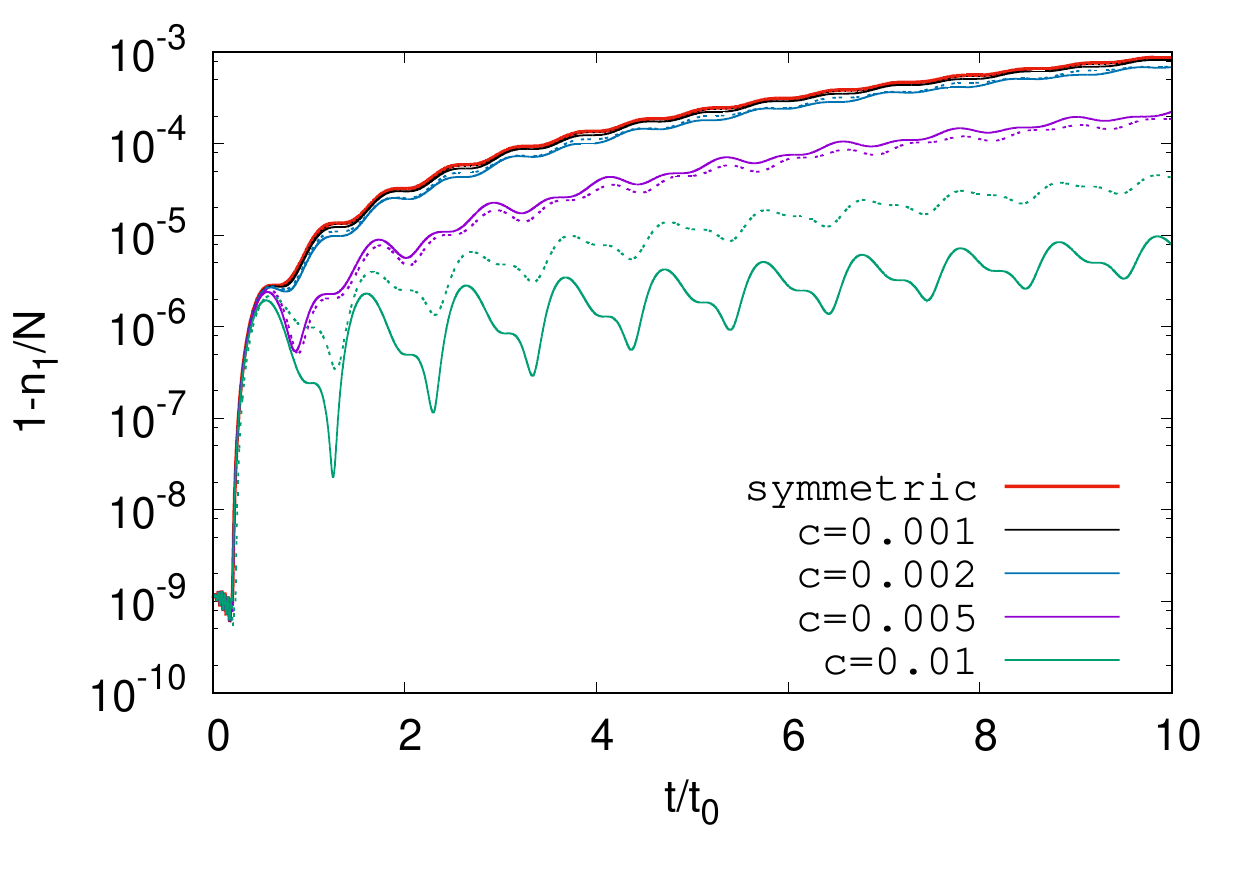}
\end{center}
\caption{ The time development of the depletion per particle $f (=\sum_{j=2}^{M} \frac{n_j}{N}=1-\frac{n_1}{N})$ of the condensate of $N=1000$ bosons in a symmetric double 
well ( curve), and an asymmetric double well trap with asymmetries $C=0.001, 0.002, 0.005$ and $0.01$. Color codes are explained in the figure itself. 
While the solid line corresponds to $f$ in the left well, the dotted line represents the same in the right well. The results shown here are computed by MCTDHB method with $M=2$ 
orbitals. See text for further details. The quantities shown are dimensionless.}
\label{fig-depletion}
\end{figure}

Having seen that a many-body calculation exhibits new features already for the time evolution of the survival probabilities at stronger $\Lambda$, 
next we would like to examine the time development of the depletion {and} fragmentation $f$ (depending on $\Lambda$) of the 
condensate which is a purely many-body quantity. 
We found the BEC to become {very slightly} depleted with time for weak interactions such as $\Lambda=0.01$. In Fig.~\ref{fig-depletion} 
the time development of the depletion $f$ of a BEC made of $N=1000$ bosons is shown for different asymmetries $C$. 
We observe that, for all values of $C$ considered here, $f$ is extremely small and therefore the system is practically condensed. Explicitly, $n_1 > 999.999$ for the 
times shown in Fig.~\ref{fig-depletion}. Thus for {such interaction strengths},
{the density per particle} of the system is accurately described by the mean-field theory. Even then, time development of $f$ exhibits some interesting features. First, the depletion is found to be maximum for the symmetric double well 
($C=0$) and gradually decreases with increasing $C$. Further, we observe an {interesting} difference between the time development of $f$ for the left and right wells. Whereas for very small $C$ the respective time evolutions of $f$ are essentially same, the difference starts to develop with $C$ and, while for 
$C \le 0.005$, the depletion for the left well is larger than that for the right well, the situation reverses for $C = 0.01$. 

%Figure 4 

\begin{figure}[!ht]
\begin{center}
\begin{tabular}{cc}
\includegraphics[width=0.50\linewidth]{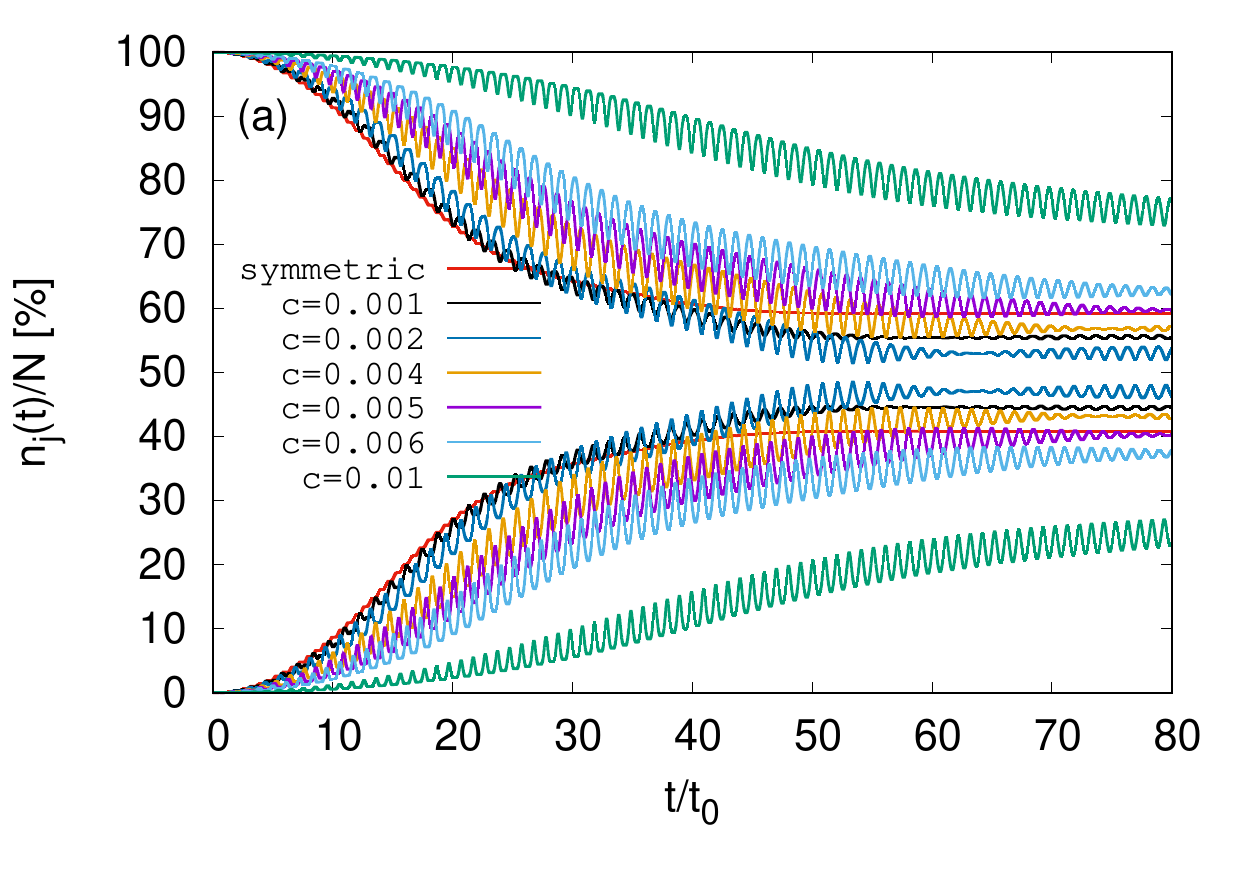} & 
\includegraphics[width=0.50\linewidth]{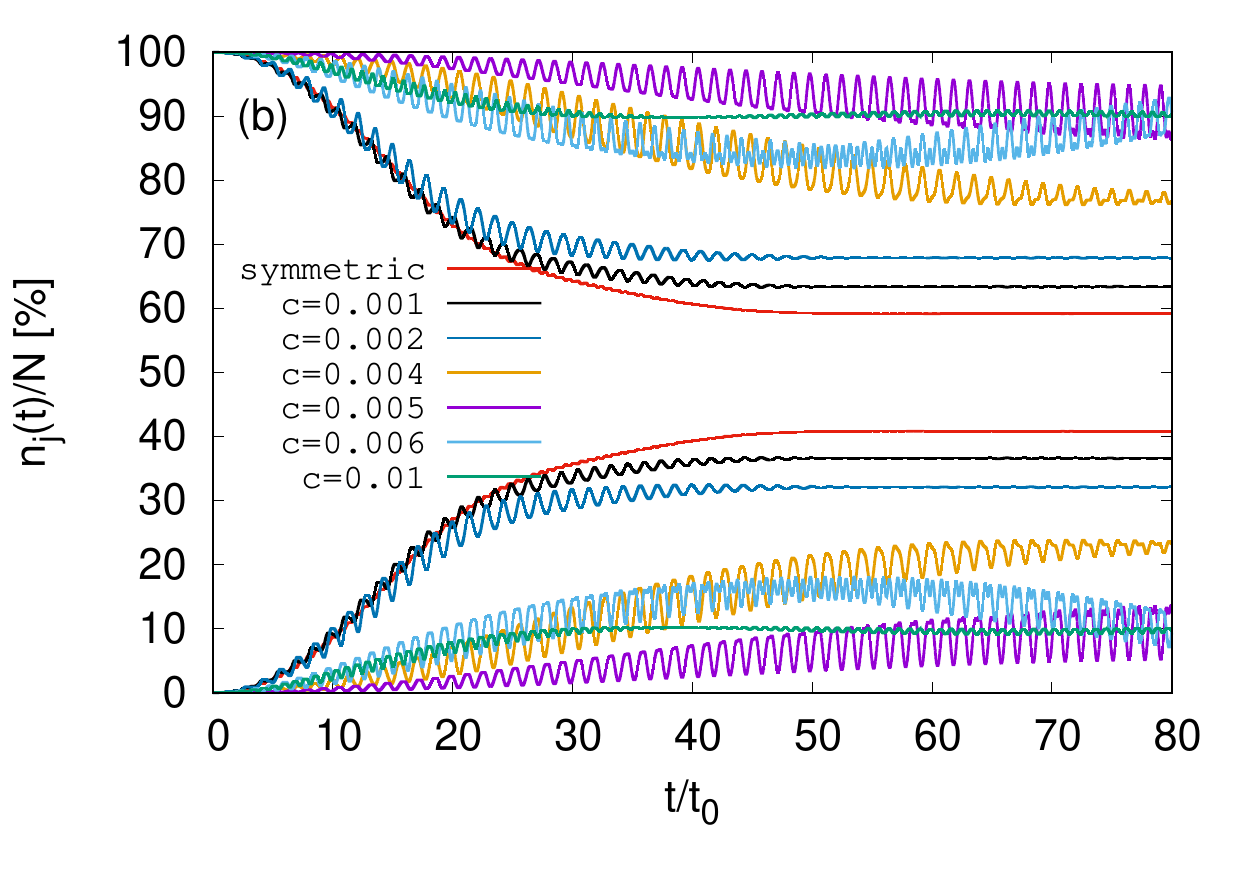}  \\
\end{tabular}
\end{center}
\caption{Fragmentation per particle $f (= \frac{n_j}{M})$ of a system of $N=1000$ bosons in an asymmetric double 
well with asymmetries $C=0.001, 0.002, 0.004, 0.005, 0.006$, and, $0.01$ for interaction 
parameter $\Lambda=0.1$ as a function of time $t$. Color codes are explained in each panel. Also the corresponding result for a symmetric double well is shown in each panel 
as a reference. The fragmentation per particle $f$ of the system for preparing the initial condensate in the left well is shown in the panel (a) while the corresponding results for 
preparing the initial state in the right well is shown in panel (b). The results shown here are obtained with the MCTDHB method with $M=2$ orbitals. 
The quantities shown are dimensionless.}
\label{fig-frag-asymm}
\end{figure}

{Next we consider a stronger interaction $\Lambda=0.1$. In Fig.~\ref{fig-frag-asymm}(a) and (b), we plot the natural occupations $\frac{n_j}{N}$ for a system of 
$N=1000$ interacting bosons as a function of time
for different asymmetries $C$. We also plot the corresponding results for the symmetric double well ($C=0$) in both panels as a reference.
The results presented here are obtained with $M=2$ orbitals. 
For all cases, we observe 
that starting from $\frac{n_1}{N}\approx 1$, the occupation in the first orbital $\frac{n_1}{N}$ gradually decreases with time. 
Simultaneously, the occupation in the second orbital $\frac{n_2}{N}$ slowly increases with time starting from a negligibly small value. 
Thus, with time the condensate becomes fragmented with a fragmentation fraction $f=\frac{n_2}{N}$. Finally, as the density 
oscillations collapse [see Fig.~\ref{fig-pl-lp1}(b)], $f$ reaches a plateau at $f=f_{col}$. Moreover, we see small oscillations in $f$ prior to attaining 
the plateau. Such oscillations are the signatures in $f$ of the time-dependent
density oscillations. As the density oscillations collapse by the time $f$ reaches the plateau, the oscillations in 
$f$ are also heavily damped and remain so at the plateau.}

Comparing the results for different asymmetries $C$, we find that both the growth rate of $f$ and $f_{col}$ depend on $C$. 
However, there is a crucial difference between the 
cases when the initial BEC state is prepared in the left [Fig.~\ref{fig-frag-asymm}(a)] and the right wells
[Fig.~\ref{fig-frag-asymm}(b)]. For the left well, $f_{col}$ first increases with increasing $C$ and 
the condensate becomes more fragmented 
with reference to the symmetric double well until $C=0.002$. With further increase in $C$, $f_{col}$ decreases and the condensate becomes less fragmented. 
On the other hand, for the right well, $f_{col}$ is found to decreases monotonically with increasing $C$, as far as $C \le 0.005$.

We may understand these findings qualitatively by treating the small asymmetry as a perturbation. 
In \cite{Sakmann2014} for each eigenstate $\vert E_n \rangle$ of the Bose-Hubbard dimer, 
its fragmentation $f_n$ as a function of the eigenstate energy per particle $E_n/N$ has been discussed. 
It has been shown both analytically and numerically that $f_n$ first increases with $E_n/N$, reaches
a maximum of $50\%$, and then decreases with further increase of $E_n/N$. For a small perturbation, such 
qualitative functional dependence is expected to remain valid. Comparing the results in
Fig.~\ref{fig-frag-asymm} with Fig.~3 of \cite{Sakmann2014}, we can infer that, for the 
parameters used in this work, the initial state for the symmetric double well lies 
on the upper part of the right-hand-branch of the $f_n$ vs $E_n/N$ curve (Fig.~3 of \cite{Sakmann2014}).  
Now, the introduction of a small asymmetry $C$ pulls down the energy for the left well and pushes up for the energy for the right well. 
Accordingly, $f_{col}$ for the left well initially increases with $C$ and then, with further increase of $C$ and consequent decrease of the initial state energy,
it crosses to the left branch of the $f_n$ vs $E_n/N$ curve (see Fig.~3 of \cite{Sakmann2014}) and starts to decrease. On {the} other hand, with the
increase in $C$ the initial eigenstate energy 
for the right well monotonically increases resulting in a monotonic decrease of $f_{col}$ (for not too large $C$). 
One would need to go beyond such a perturbation-based analysis to understand the behavior of fragmentation for $C>0.005$ in the right well, see Fig.~\ref{fig-frag-asymm}(b).

%Figure 5
\subsubsection{Many-particle position and momentum variance}
\label{variance}
\begin{figure}[!ht]
\begin{center}
\includegraphics[width=0.45\linewidth]{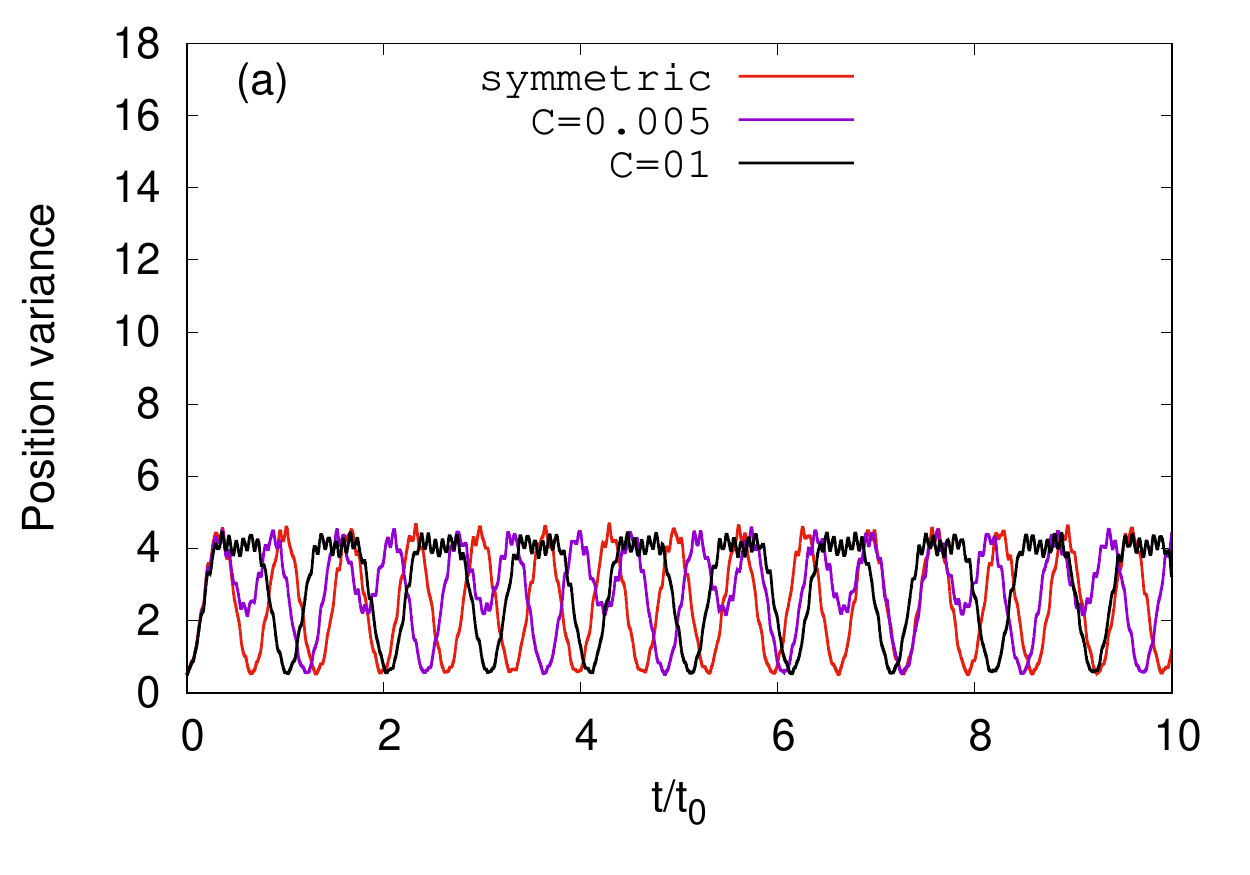} 
\includegraphics[width=0.45\linewidth]{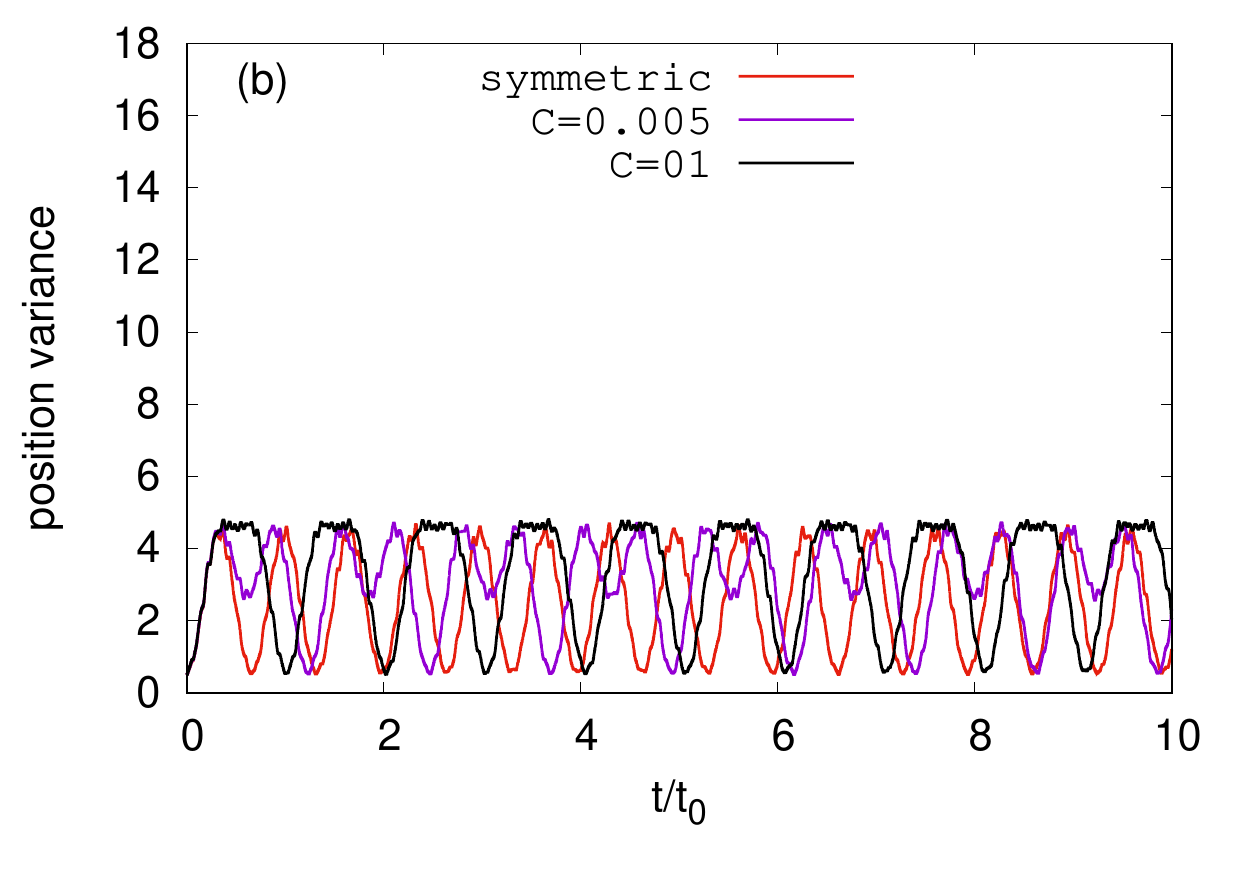}  \\
\includegraphics[width=0.45\linewidth]{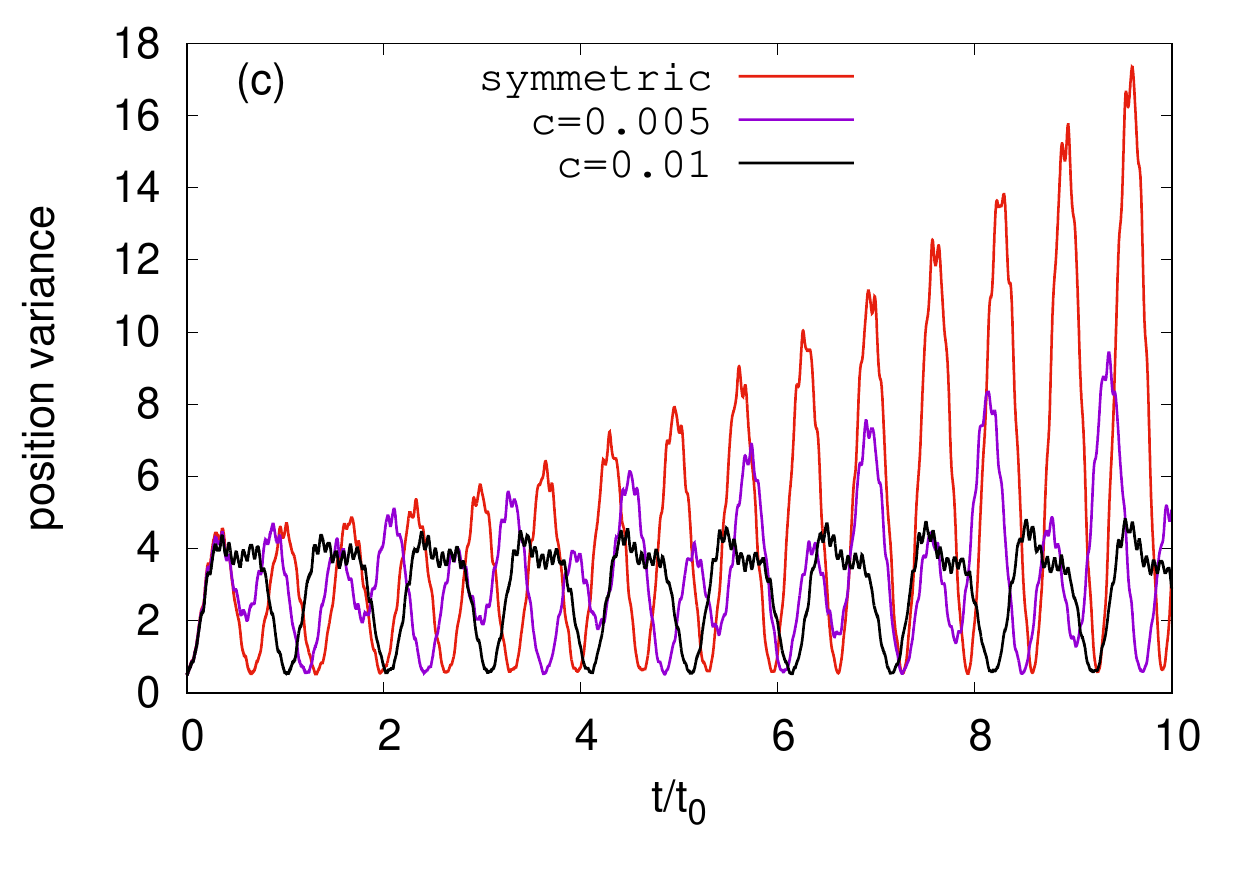} 
\includegraphics[width=0.45\linewidth]{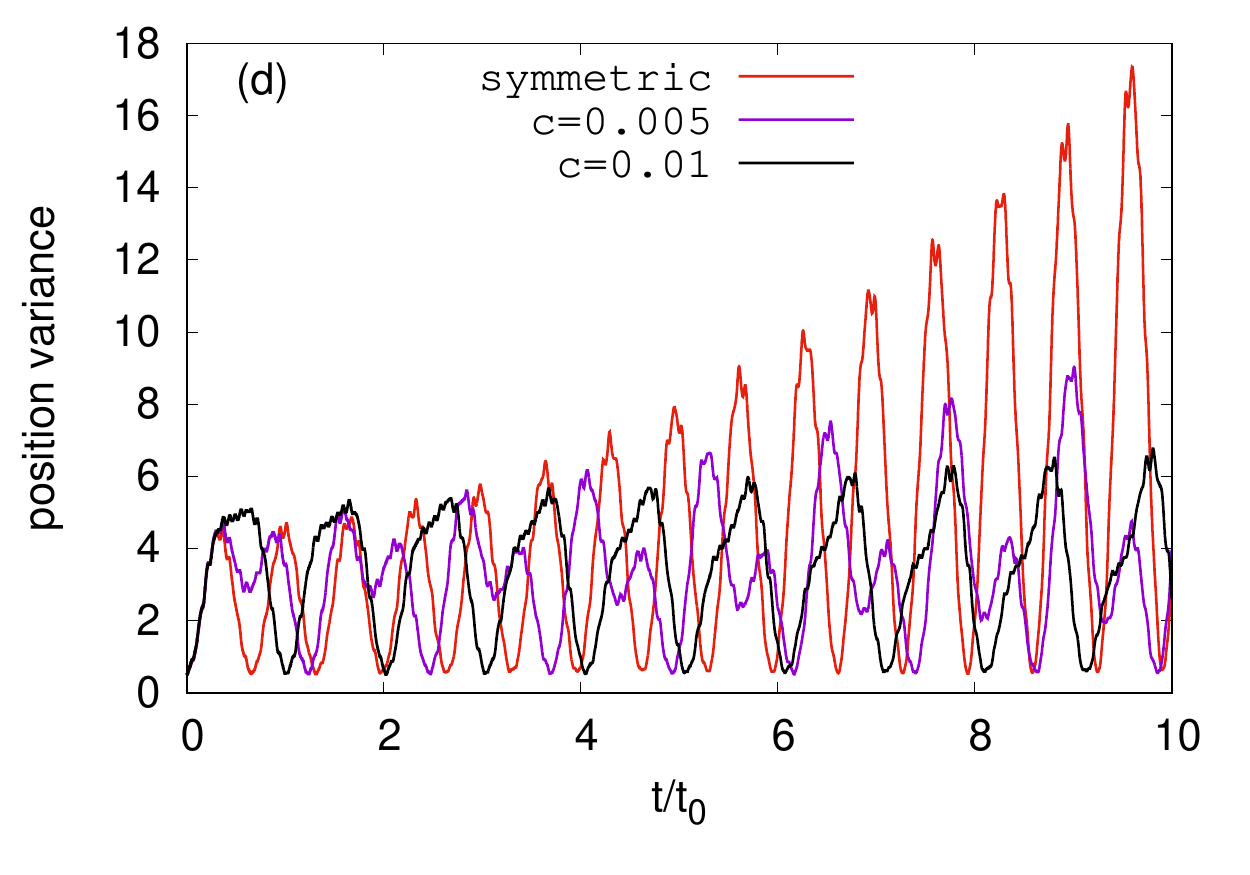}  \\
\end{center}
\caption{many-particle position variance per particle $\frac{1}{N} \Delta^2_{\hat X}$  of a system 
of $N=1000$ bosons in a symmetric $(C=0)$ and an asymmetric double well with asymmetries $C=0.005$ and $0.01$ for interaction parameter $\Lambda=0.01$ as a function of time $t$. 
Mean-filed $\frac{1}{N} \Delta^2_{\hat X}$ of the system for preparing the initial condensate state in the left and the right wells are shown in panel (a) and (b), 
respectively. Corresponding MCTDHB results with $M=2$ orbitals for $\frac{1}{N} \Delta^2_{\hat X}$ for preparing the initial condensate state in the right well are 
exhibited in panel (b) and (d), respectively. Color codes are explained in each panel. The quantities shown are dimensionless.}
\label{fig-variance}
\end{figure}

{Next, we consider the time evolution of the many-particle position and momentum variances. {These quantities characterize the fluctuations in the particles' positions and momenta in the junction. Although not easy to measure, they are fundamental quantum mechanical observables.}
%Here we discuss the variance of the many-particle operator. 
%For a corresponding discussion on the variance of the many-particle momentum operator, we refer to the Appendix~\ref{J}.
%which can deviate from their corresponding mean-field results even when the mean-field theory is expected to accurately describe the density per particle of the system, 
Since these quantities depend on the actual number of depleted or fragmented atoms, it is expected that prominent signatures of the depletion 
and fragmentation of the condensate would show up in these variables. For $\Lambda=0.01$, it has been shown above that the system remains practically condensed for a long time 
(see above) and therefore, its out-of-equilibrium dynamics should be adequately described by the mean-field theory. So, first we study the time evolution of the many-particle position
variance $\frac{1}{N} \Delta^2_{\hat X}$ at the mean-field level.
%Accordingly, we next study the variance of the many-particle position and momentum operators. 
We present our results for different asymmetries $C$ and a fixed $\Lambda=0.01$ in Fig.~\ref{fig-variance}(a)
%we plot the many-particle position variance $\frac{1}{N} \Delta^2_{\hat X}$ {of the system for preparing the initial BEC state in the left well
and (b) for preparing the initial BEC state in $V_L(x)$ and $V_R(x)$, respectively. 
For comparison, we also plot $\frac{1}{N} \Delta^2_{\hat X}$ for the symmetric double well in both panels.} 

{We observe that for both $V_L(x)$ and $V_R(x)$ of the asymmetric double well trap, $\frac{1}{N} \Delta^2_{\hat X}$ oscillates with a frequency which equals to the Rabi frequency. This is in contrast to the case of the symmetric double well in which $\frac{1}{N} \Delta^2_{\hat X}$ oscillates with a frequency equal to twice the Rabi frequency.  This is due to the incomplete tunneling between the two wells of the asymmetric double well trap, and that there is always a remnant in the each well which is further manifested in the irregular peaks of oscillations of $\frac{1}{N} \Delta^2_{\hat X}$. We observe that with increase in $C$ starting from the symmetric double well ($C=0$), the peaks of the oscillations of $\frac{1}{N} \Delta^2_{\hat X}$ first split into two sub-peaks which gradually turn into broad peaks for $C=0.01$.  Further, we observe high-frequency small-amplitude oscillations on top of the peaks of the large-amplitude oscillations. Such high-frequency oscillations are because of the 
breathing-mode oscillations of the system in the asymmetric double well and can be seen more vividly in the many-particle momentum variance (see below). Also, the minima of the oscillations are slightly higher than $0.5$ for all times $t > 0$. Moreover, comparing the panels (a) and (b), we see that the peak values of the oscillations are slightly higher for the right well $V_R(x)$.} {All of these quantify the fluctuations in the particles' positions in the asymmetric double well at the mean-field level.}

{As discussed in Sec.~\ref{dynamics}, the many-particle position variance can deviate from their corresponding mean-field results even when the mean-field theory is expected to accurately describe the density per particle of the system. So, we now study the time evolution of the many-particle position variance $\frac{1}{N} \Delta^2_{\hat X}$ at the many-body level.}  For all cases, $\frac{1}{N} \Delta^2_{\hat X}$ is found to grow in an oscillatory manner. For the symmetric double well ($C=0$), the maxima of 
$\frac{1}{N} \Delta^2_{\hat X}$ grow approximately quadratically, also see~\cite{Klaiman2016}. This growth is slower for an asymmetric double well, where
the growth rate decreases with increasing $C$.
This is consistent with our earlier observation that the depletion of the condensate is maximal in a symmetric double well. 
{As observed at the mean-field level, here also, the oscillations of $\frac{1}{N} \Delta^2_{\hat X}$ are irregular in nature. However, now the two sub-peaks are of unequal heights and the difference between them grows with time $t$ for both $V_L(x)$ and $V_R(x)$. Comparison between the left [panel (c)] and the right wells [panel (d)] shows that, while the higher sub-peaks is on the left side for $V_L(x)$, it appears on the right side for $V_R(x)$. Further}, while for $C = 0.005$ the maximal values for the $\frac{1}{N} \Delta^2_{\hat X}$ in left well 
are larger than those in the right well, the situation reverses for $C=0.01$. This is again consistent with our earlier observation (in Fig.~\ref{fig-depletion}) that the system 
in the left well is more depleted until $C=0.005$, whereas the system in the right well is more depleted for $C=0.01$.
%Figure 6

\begin{figure}[!ht]
\begin{center}
\includegraphics[width=0.45\linewidth]{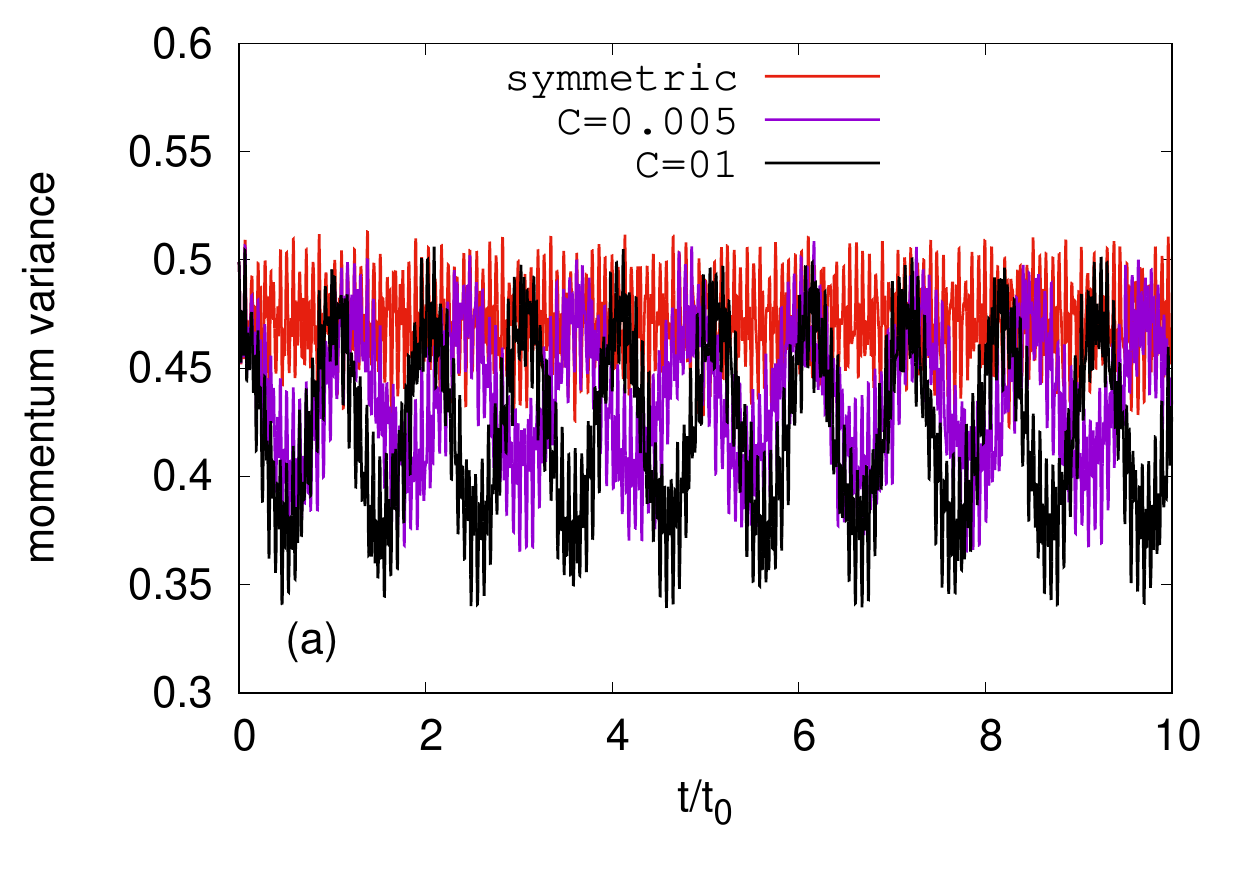} 
\includegraphics[width=0.45\linewidth]{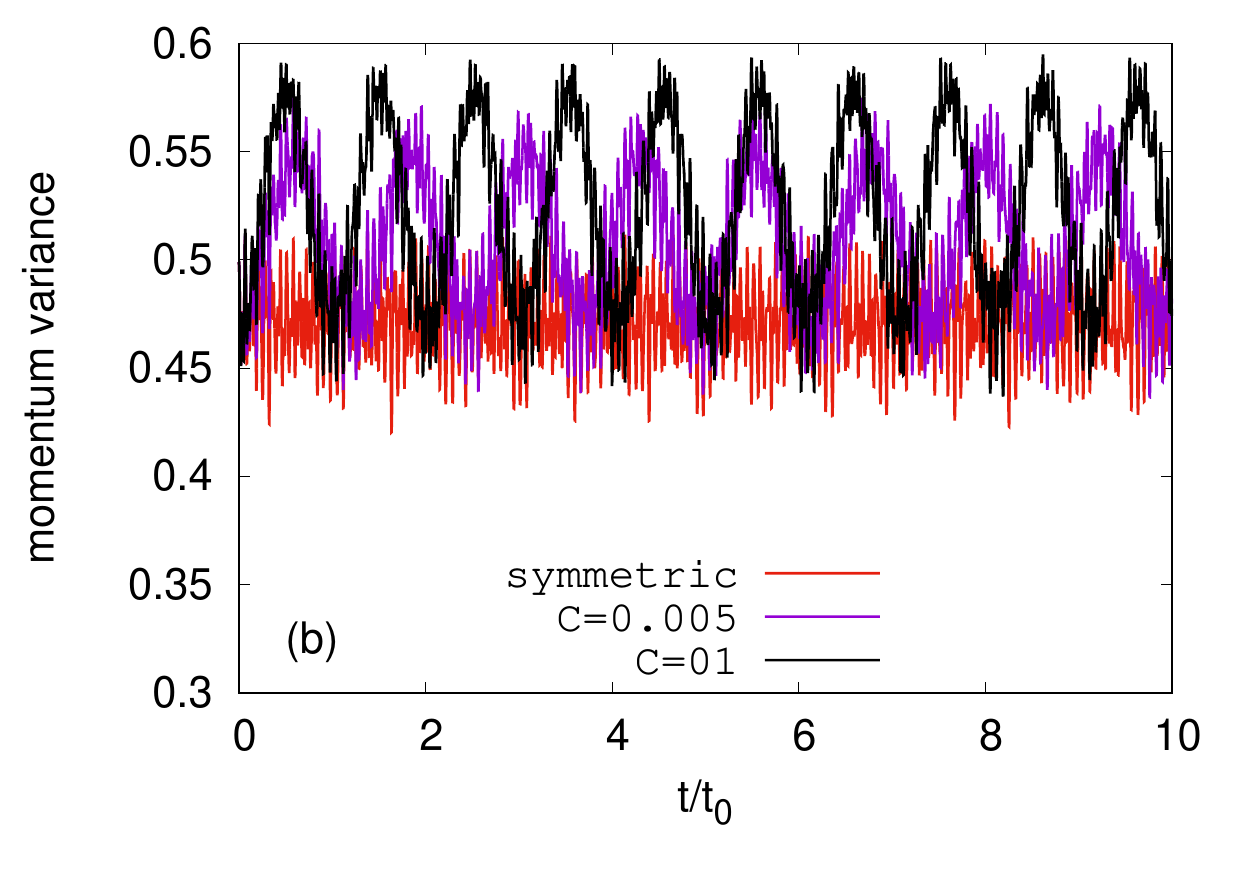} \\
\includegraphics[width=0.45\linewidth]{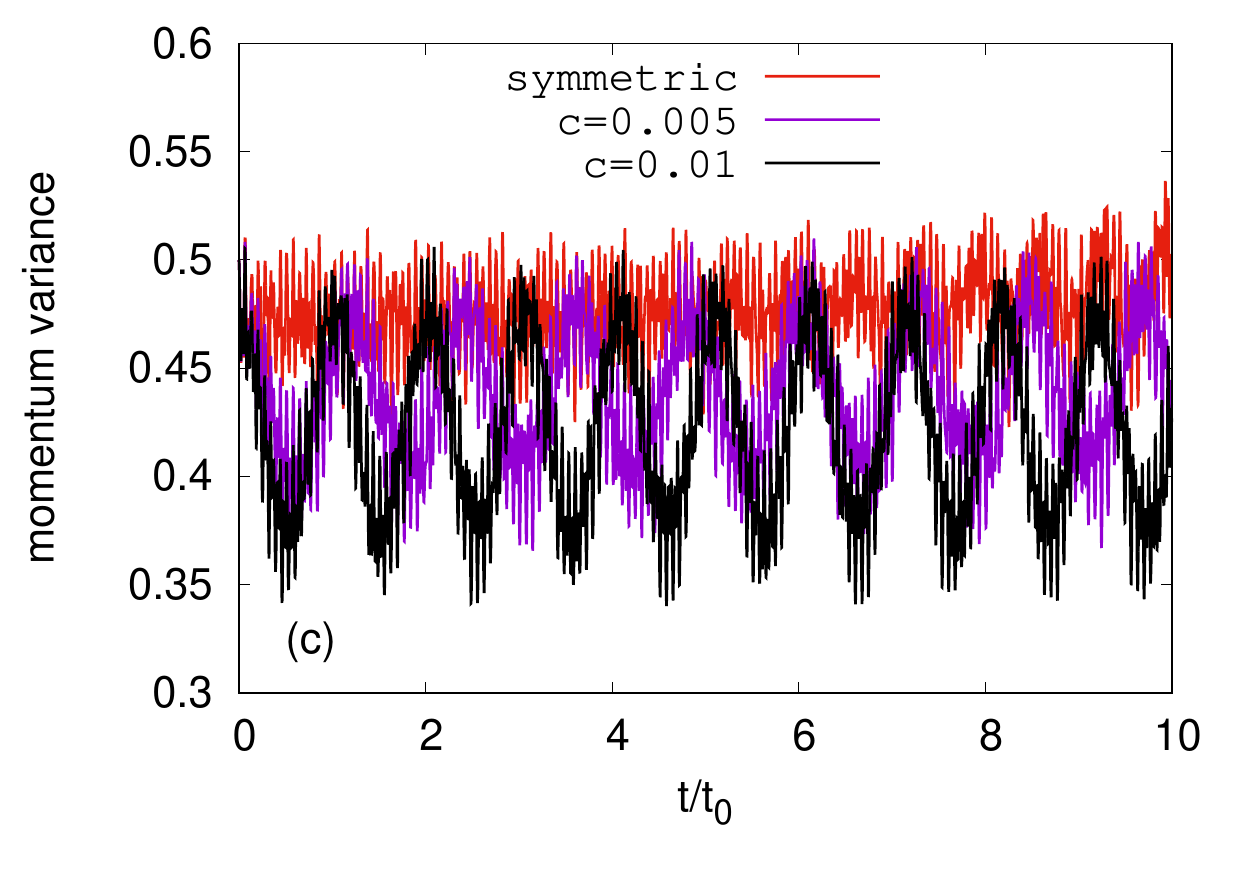} 
\includegraphics[width=0.45\linewidth]{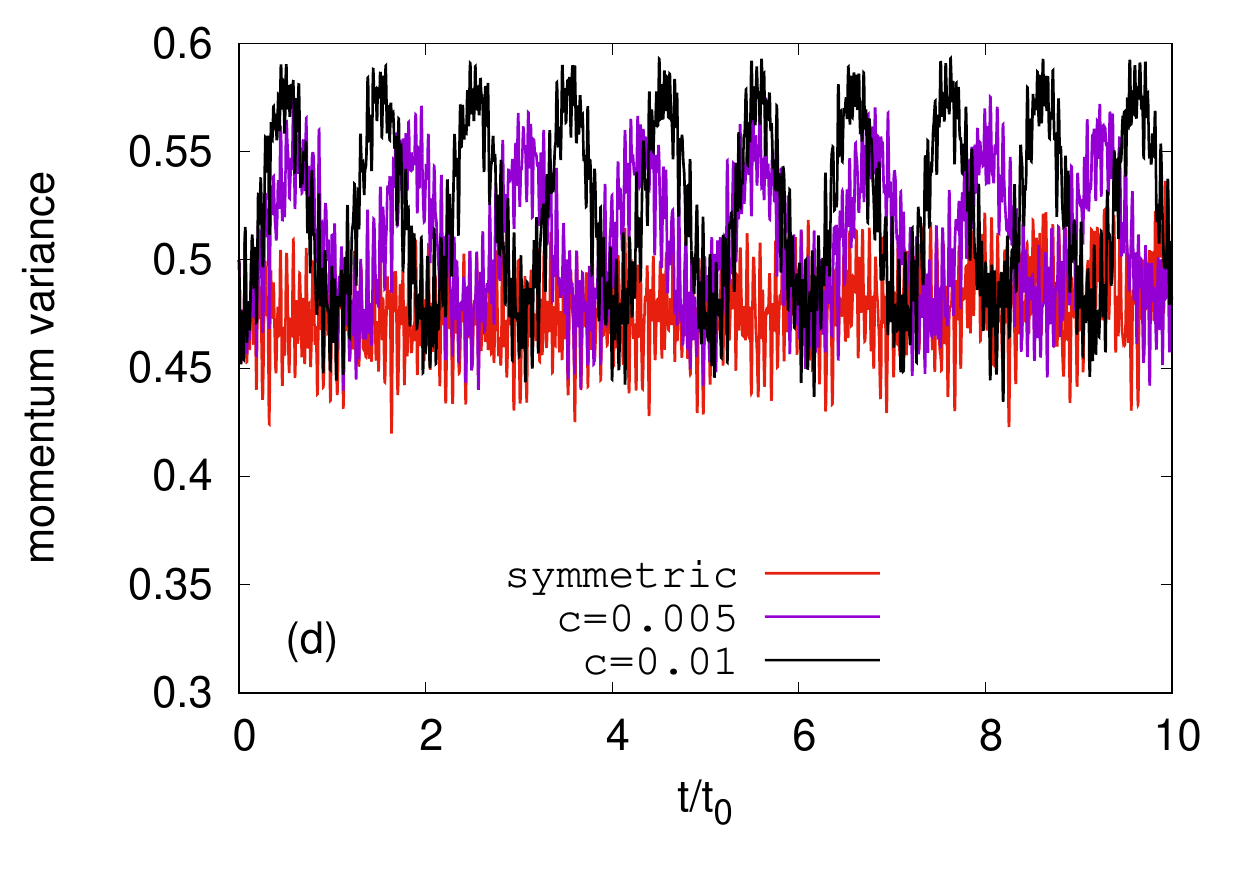}  \\ 
\end{center}
\caption{many-particle momentum variance per particle $\frac{1}{N} \Delta^2_{\hat P}$ of a system of of $N=1000$ bosons in a symmetric $(C=0)$ and an asymmetric double well 
with asymmetries $C=0.005$ and $0.01$ for interaction parameter $\Lambda=0.01$ as a function of time $t$.Mean-filed $\frac{1}{N} \Delta^2_{\hat P}$ 
of the system for preparing the initial condensate state in the left and the right wells are shown in panel (a) and (b), 
respectively. Corresponding MCTDHB results with $M=2$ orbitals for $\frac{1}{N} \Delta^2_{\hat P}$ for preparing the initial condensate state in the right well are 
exhibited in panel (b) and (d), respectively. Color codes are explained in each panel. The quantities shown are dimensionless.}
\label{fig-mom-variance}
\end{figure}

Next, in Fig.~\ref{fig-mom-variance}, we plot the many-particle momentum variance $\frac{1}{N} \Delta^2_{\hat P}$ of the system for starting 
the dynamics from $V_L(x)$ and $V_R(x)$, respectively. {We studied $\frac{1}{N} \Delta^2_{\hat P}$ both at the mean-field and the many-body levels. In Fig.~\ref{fig-mom-variance}(a) and (b) we present the mean-field results of $\frac{1}{N} \Delta^2_{\hat P}$ for the left and the right wells, respectively.}
In each panel, we also plot $\frac{1}{N} \Delta^2_{\hat P}$ for the symmetric double well {for comparison. For all cases, we} observe
two oscillations associated with the time evolution of $\frac{1}{N} \Delta^2_{\hat P}$: The first, with a larger amplitude and frequency equal to twice the Rabi frequency and,  the second, with a smaller amplitude but a higher frequency. The first one is a manifestation of the density oscillations, whereas the second one is due to the breathing oscillations of the system.  {However, while in the symmetric double well the amplitude of the breathing mode oscillations are larger than those of the density oscillations, the situation is reversed in the asymmetric double well. }

 {A closer examination} {of the high frequency breathing mode oscillations suggests that these may arise due to the transition of two bosons from the lowest energy band to the second band or one boson from the lowest band to the third band. An analysis by a  linear-response theory in the line of Ref.~\cite{Marcus} is required to attribute such high-frequency oscillations to a particular transitions unambiguously and accurately. In any case, it can be safely said that one needs to consider higher bands to take into account such high-frequency breathing mode oscillations and} {this, of course, is beyond the scope of the standard Bose-Hubbard model.}
Moreover, though 
$\frac{1}{N} \Delta^2_{\hat P}$ starts from the same value for both the left and right well, there is a $\Pi-$phase difference 
between the oscillations for the left well and  the right well. Explicitly, the momentum variance first decreases when starting from the lower (left) well, 
whereas it first increases when starting from the higher (right) well. This can be understood from an energetic point of view. The BEC tunneling from the lower
to upper well initially loses kinetic energy (momentum) and gains kinetic energy when tunneling from the higher to lower well. The momentum variance behaves accordingly. 

{Finally, we show the corresponding MCTDHB results of $\frac{1}{N} \Delta^2_{\hat P}$ with $M=2$ orbitals in Fig.~\ref{fig-mom-variance}(c) and (d) for the left and the right wells, respectively. We find that the MCTDHB dynamics of $\frac{1}{N} \Delta^2_{\hat P}$ is similar to the corresponding mean-field dynamics. Actually, the many-particle momentum variance depends on the derivatives of the orbitals. For a weak asymmetry and a weakly interacting system, the shape of the orbitals deviate only slightly from their corresponding (non-interacting and) mean-field shape, and this leads to even smaller derivatives thereby producing practically same $\frac{1}{N} \Delta^2_{\hat P}$ both at the mean-field and the many-body levels.}
  
\subsection{Universality of the fragmentation dynamics in an asymmetric double well}
\label{universality}

A unique many-body feature {predicted} in the dynamics of BECs in a {symmetric double well 
is the universality of the degree of fragmentation with respect to $N$ for a fixed $\Lambda$~\cite{Sakmann2014}. 
It was first established by solving the many-body Schr\"odinger equation and then  
using the Bose-Hubbard dimer, it was also shown that the universality of fragmentation in a symmetric double well is a general many-body phenomenon~\cite{Sakmann2014}}. 
Also, in the previous subsection, we have already found a significant effect of the asymmetry of the trap on the 
time evolution of the survival probabilities, fragmentation, and 
the many-particle position and momentum variances of BEC in an asymmetric double well trap.
Naturally, questions arise if the universality of the fragmentation exists in an asymmetric trap
and if so, how it is affected by the asymmetry of the trap.

Once again we start with the corresponding symmetric double well as a reference. In Fig.~\ref{fig-universality}(a) we have plotted the natural occupations for different 
$N$ keeping $\Lambda$ fixed. As discussed above, we see that initially only one natural orbital is occupied with $\frac{n_1}{N} \approx 1$ 
and negligibly small $f$ for all cases shown in Fig.~\ref{fig-universality}. 
However, with time the 
second natural orbital starts to be occupied, the system becomes fragmented and, during the collapse of the density oscillations, 
the occupations of the natural orbitals reach the 
same plateau for different numbers of bosons $N$ keeping $\Lambda$ fixed.  
The values at the plateau are about $\frac{n_1}{N}=60\%$ and $\frac{n_2}{N}=40\%$, respectively. 
Hence for all cases, after the collapse of the density oscillations the system becomes $f_{col} \approx 40\%$ 
fragmented irrespective of $N$, showing a universal fragmentation dynamics~\cite{Sakmann2014}. 

Next, we consider an asymmetric double well with a very small asymmetry $C=0.001$. Fig.~\ref{fig-universality}(b) shows the results 
for $V_L(x)$ and Fig.~\ref{fig-universality}(c) for $V_R(x)$. For both wells, qualitatively, we see the same dynamics as in the symmetric double well. 
We observe that following an oscillatory growth, $f$ reaches the same plateau $f_{col}$ irrespective of the number of particles $N$ for a fixed
$\Lambda$. Therefore, the universality of the fragmentation dynamics also persists in an asymmetric double well.  
However, quantitatively $f_{col}$ for the left well differs from that for the right, $f_{col} \approx 45\%$ versus $f_{col} \approx 35\%$, respectively. 
Here, an interesting point is that $f_{col}$ for the symmetric well is actually the mean of $f_{col}$ for the two wells of the asymmetric double well. 
As shown in Ref~\cite{Sakmann2014}, the fragmentation 
depends on the energy per particle. Since we have introduced the asymmetry by adding a linear slope of a fixed gradient, 
it pushes up the right well by about the same amount as it pulls down the left well. 
Therefore, the changes in $f_{col}$ for both wells are expected to be similar but in opposite directions, leading to the above relation between the fragmentation values. 

{As discussed earlier, the many-particle position variance $\frac{1}{N} \Delta^2_{\hat X}$ bears prominent signatures of the fragmentation. 
Hence, next, we study the time evolution of the many-particle position variance $\frac{1}{N} \Delta^2_{\hat X}$, {both at the mean-field and the many-body levels,} to explore the possible manifestation of the universality of the 
fragmentation dynamics. {In Fig.~\ref{fig-universality-var} we plot the MCTDHB results with $M=2$ orbitals of $\frac{1}{N} \Delta^2_{\hat X}$, for different $N$ but the same $\Lambda$, as a function of time for starting the dynamics from 
both the left [panel (a)] and the right [panel (b)] wells. We also plot the corresponding mean-field results in both panels for comparison.  In the mean-field theory, there is only one parameter $\Lambda$ and therefore, we have only one curve for the time development of $\frac{1}{N} \Delta^2_{\hat X}$ for a particular $\Lambda$ irrespective of $N$. On the other hand, at the many-body level, we find different time development for $\frac{1}{N} \Delta^2_{\hat X}$ for different $N$ corresponding to the same $\Lambda$.  For all $N$ corresponding to the same $\Lambda$ and both wells, $\frac{1}{N} \Delta^2_{\hat X}$ exhibits an oscillatory growth before reaching a saturation at a mean value $\frac{1}{N} \Delta^2_{\hat X}\rvert_{sat}$. While the growth rate of $\frac{1}{N} \Delta^2_{\hat X}$ for different $N$ corresponding to a fixed $\Lambda$ are the same, 
%We observe that at the mean-field level, $\frac{1}{N} \Delta^2_{\hat X}$ smoothly oscillates with a frequency equal to twice the Rabi frequency between a minima $\sim 0.5$ and a maxima $\sim 4.5$ for both wells. This implies that for such a small asymmetry with $C=0.001$, the density of the system practically performs full oscillations at the mean-field level for both $V_L(x)$ and $V_R(x)$. On the other hand, at the many-body level, we find for both wells that $\frac{1}{N} \Delta^2_{\hat X}$ exhibits an oscillatory growth before reaching a saturation at a mean value $\frac{1}{N} \Delta^2_{\hat X}\rvert_{sat}$. While for all $N$, the frequency of oscillations of $\frac{1}{N} \Delta^2_{\hat X}$, prior to reaching the saturation, is twice the Rabi frequency, as observed at the mean-field level,} 
the saturation values $\frac{1}{N} \Delta^2_{\hat X}\rvert_{sat}$ increase with $N$. Moreover, we note that the time required 
to reach the saturation and the saturation values $\frac{1}{N} \Delta^2_{\hat X}\rvert_{sat}$ are similar for both the left and right wells. 
%Further, the ratio between the two saturation values $\frac{1}{N} \Delta^2_{\hat X}\rvert_{sat}$ for two different particle numbers $N_1$ and $N_2$ 
%is of the same order as the ratio of the particle number $\frac{N_1}{N_2}$, 
%viz. $\frac{\frac{1}{N} \Delta^2_{\hat X}\rvert_{sat}^{N=N_1}}{\frac{1}{N} \Delta^2_{\hat X}\rvert_{sat}^{N=N_2}} \sim \frac{N_1}{N_2}$, where
%$\frac{1}{N} \Delta^2_{\hat X}\rvert_{sat}^{N}$ is the saturation value for $N$ particles. 
Further, for both wells, the saturation value $\frac{1}{N} \Delta^2_{\hat X}\rvert_{sat}^{N=N_1}$ for a BEC made of $N=N_1$ particles is of the same order of magnitude, viz.,   
$\frac{1}{N} \Delta^2_{\hat X}\rvert_{sat}^{N=N_1} \sim N_1$. For example, 
in Fig~\ref{fig-universality-var}(a) and (b), for both wells, while $\frac{1}{N} \Delta^2_{\hat X}\rvert_{sat}^{N=100} \sim 10^{2}$ for $N=100$, it increases to 
$\frac{1}{N} \Delta^2_{\hat X}\rvert_{sat}^{N=1000} \sim 10^3$ and $\frac{1}{N} \Delta^2_{\hat X}\rvert_{sat}^{N=10000} \sim 10^4$ for $N=1000$ and $10000$, respectively.}

These observations can be understood as follows. In Fig.~\ref{fig-universality}, 
we have seen that $f_{col}$ for the left well is only about $10\%$ higher than that of the 
right well, for all $N$. Naturally, the actual occupation numbers $n_2$ are of the same order of magnitude for both wells for all $N$. 
Since $\frac{1}{N} \Delta^2_{\hat X}$ depends on the actual value of $n_2$~\cite{Klaiman2015}, 
its saturation values for a particular $N$ are of the same order of magnitude (as a power of $10$) 
for both the wells. Similarly, due to the universality of fragmentation dynamics, 
$f_{col}$ corresponding to different $N$ and same $\Lambda$ have the same value for a particular well. 
Therefore, the actual number of fragmented atoms $n_2$ increases by a factor of $\frac{N_2}{N_1}$ 
for an increase of $N$ from $N_1$ to $N_2$. Accordingly, $\frac{1}{N} \Delta^2_{\hat X}\rvert_{sat}$ 
also increases by a factor of $\frac{N_2}{N_1}$.

{
To stress this point further, in Fig.~\ref{fig-universality-var}(c) and (d), we divide $\frac{1}{N} \Delta^2_{\hat X}\rvert_{sat}$ by $N$, and
plot $\frac{1}{N^2} \Delta^2_{\hat X}$ for different $N$, keeping $\Lambda$ fixed, for both wells.
Again we find qualitatively similar time development of $\frac{1}{N^2} \Delta^2_{\hat X}$ for both wells. 
As before, for both wells, $\frac{1}{N^2} \Delta^2_{\hat X}$ also
exhibit an oscillatory growth followed by an equilibration after the collapse of the density oscillations. 
However, the important point is that the curves for different $N$, 
keeping $\Lambda$ same, saturate to the same mean value about which $\frac{1}{N^2} \Delta^2_{\hat X}$ keeps on oscillating. 
This is the signature of the universal fragmentation 
dynamics. Therefore, the universality of fragmentation is a quite robust many-body phenomena and its signature appears in all many-body quantities that depend
on the occupation numbers of the natural orbitals.}

%Figure 7
\begin{figure}[!ht]
\begin{center}
\begin{tabular}{cc}
\includegraphics[width=0.5\linewidth]{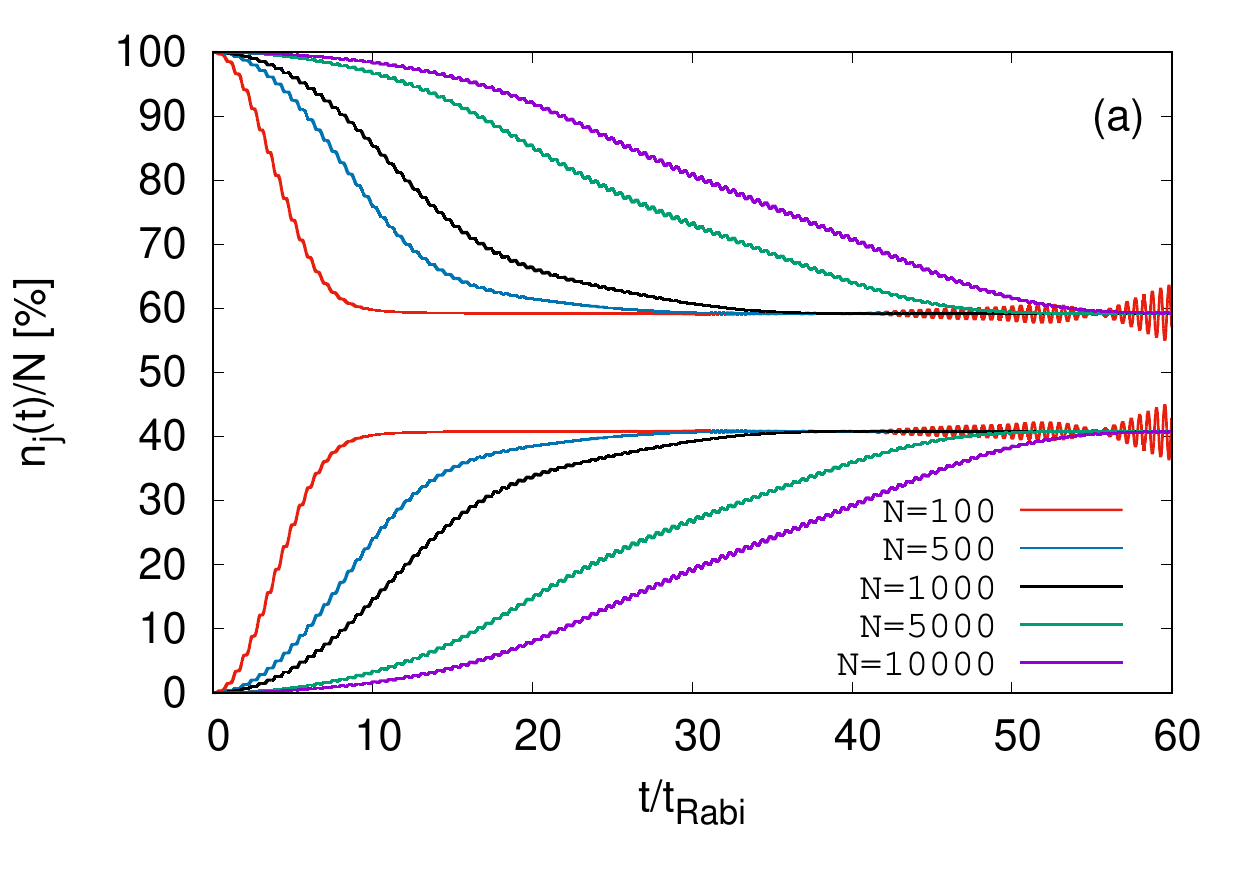} & \\
\includegraphics[width=0.5\linewidth]{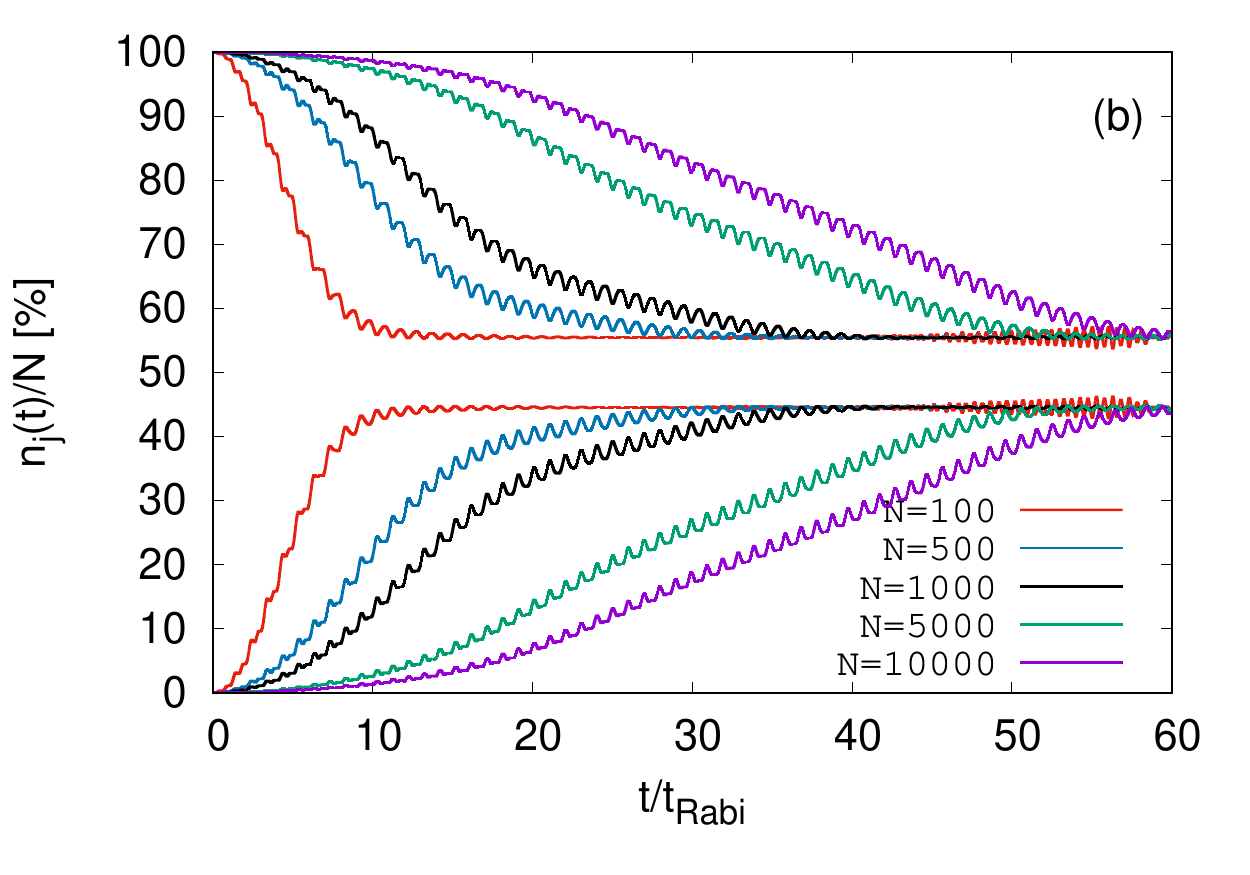} & \\
\includegraphics[width=0.5\linewidth]{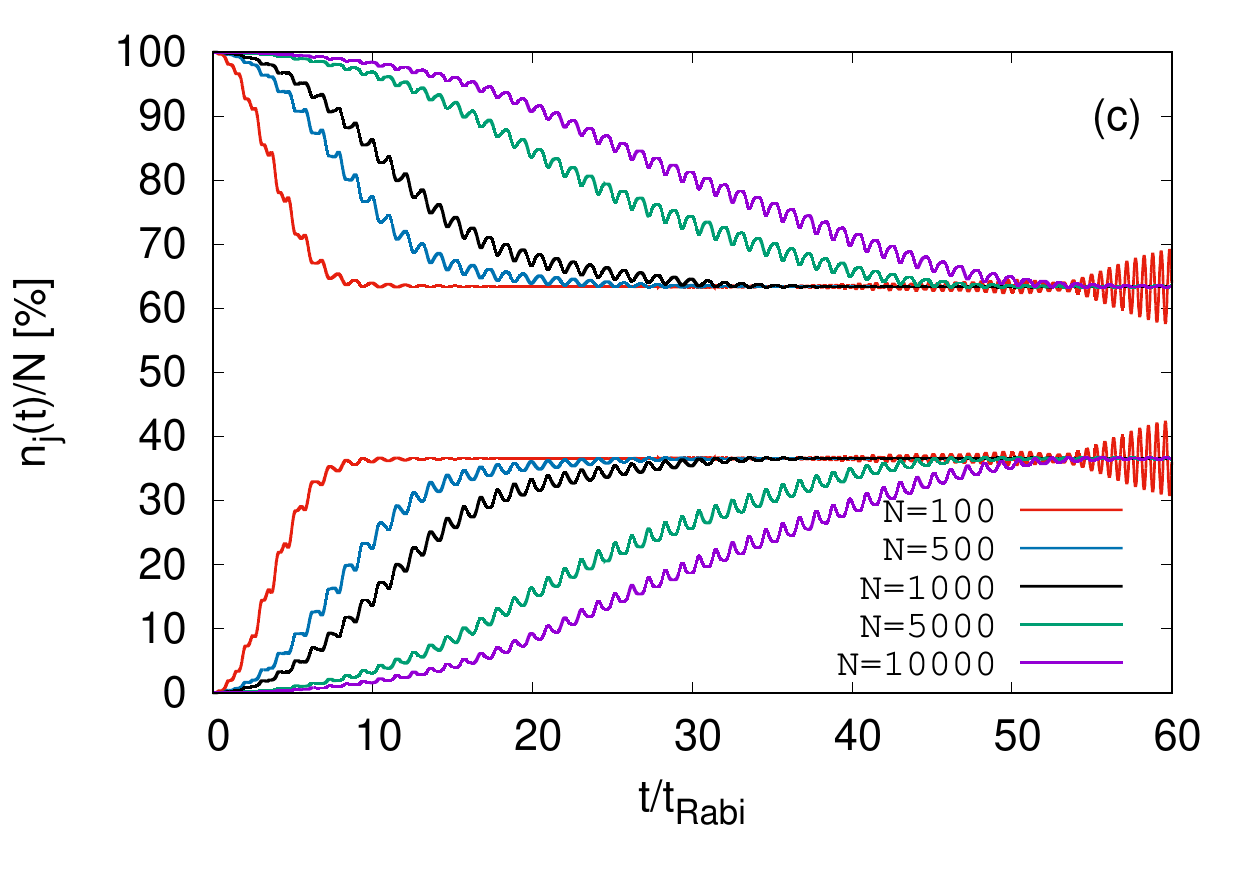} & \\
\end{tabular}
\end{center}
\vglue -0.9 truecm
\caption{Universality of the fragmentation dynamics for $\Lambda=0.1$. (a) The natural occupations $\frac{n_j}{N}$ as a 
function of $t$ for BECs consisting of different number $N$ of bosons in the 
symmetric double well. For all $N$, we prepared the initial condensed state in the left well. (b) Same as in panel (a) but for
an asymmetric double well of asymmetry $C=0.001$. (c) The corresponding time evolution of 
$\frac{n_j}{N}$ for $C=0.001$ when the initial condensate is prepared in the right well. 
In all panels, the upper curve represents $\frac{n_1}{N}$ while the lower curve shows the corresponding 
$\frac{n_2}{N}$. All the $\frac{n_j}{N}$ shown here are computed by the MCTDHB method with $M=2$ orbitals. For further details, 
refer to the text. Color codes are explained in each panel. The quantities shown are dimensionless.}
\label{fig-universality}
\end{figure}

%Figure 8 
\begin{figure}[!ht]
\begin{center}
\begin{tabular}{cc}
\includegraphics[width=0.5\linewidth]{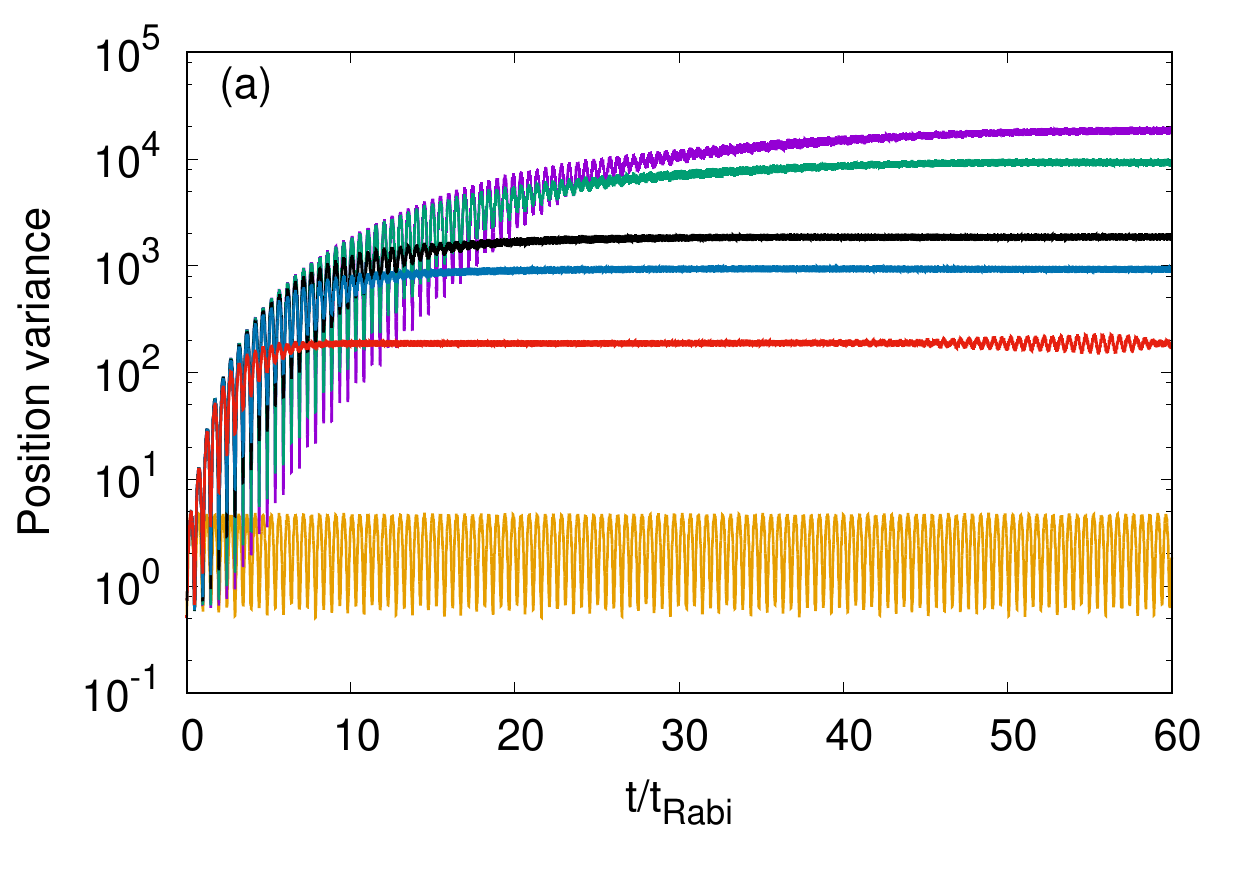} & 
\includegraphics[width=0.5\linewidth]{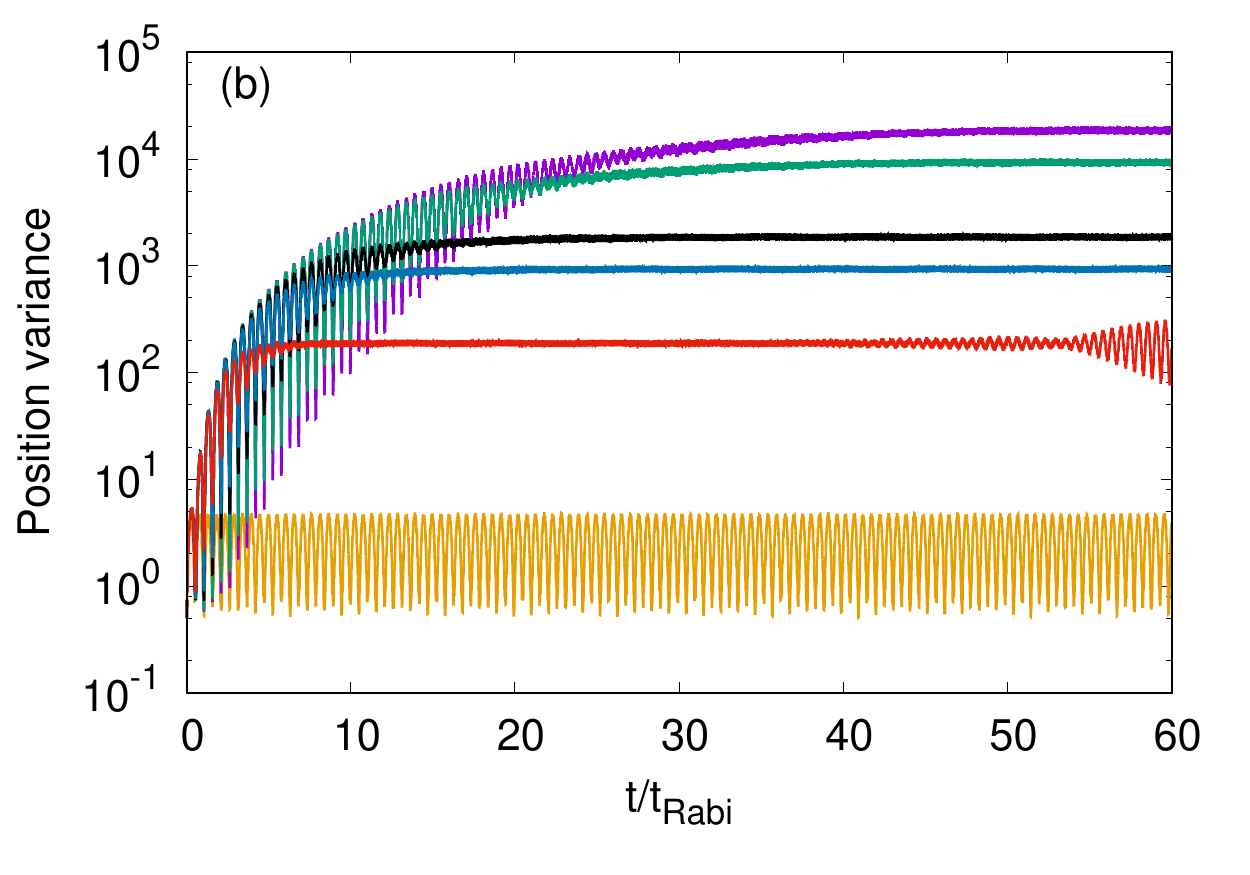}  \\
\includegraphics[width=0.5\linewidth]{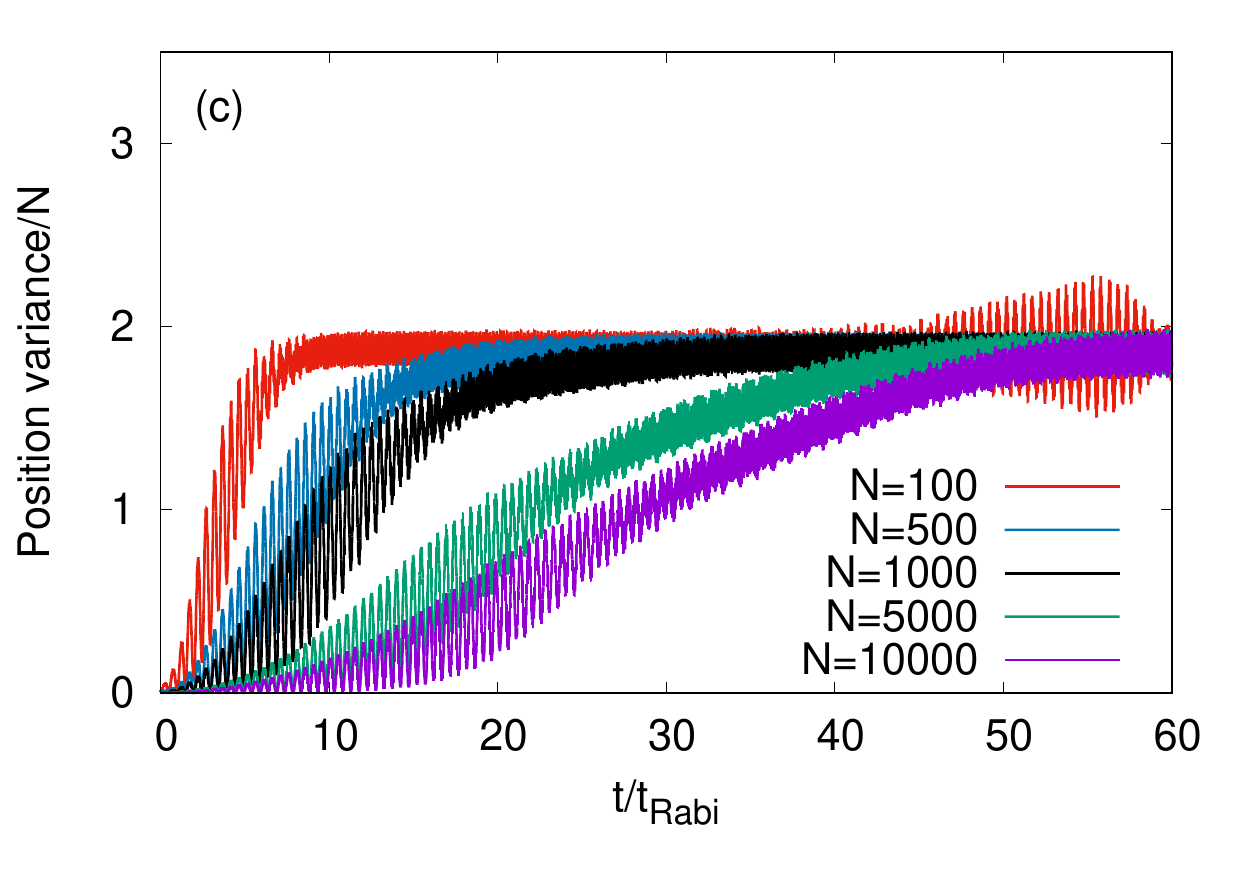} & 
\includegraphics[width=0.5\linewidth]{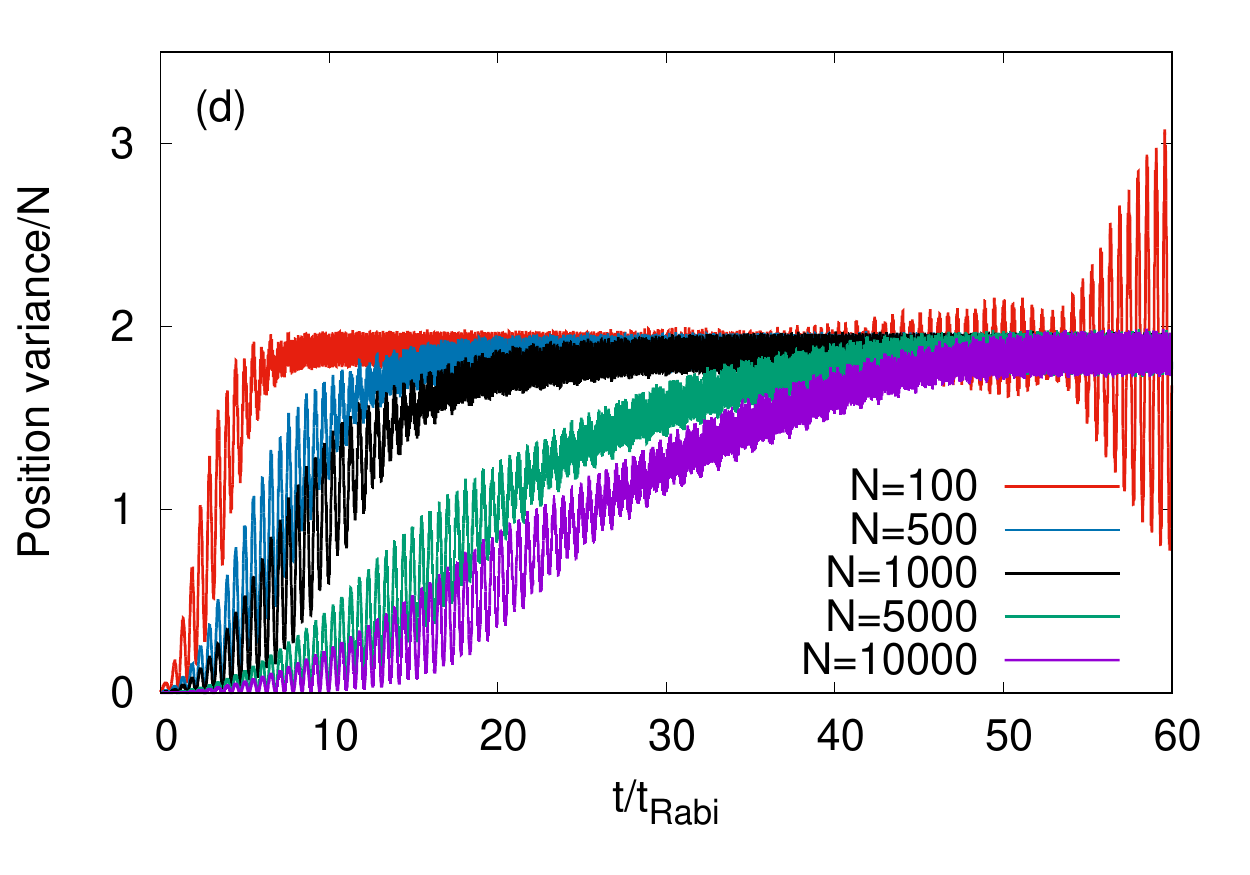}  \\
\end{tabular}
\end{center}
\caption{Signature of universality of the fragmentation dynamics, as shown in Fig.~\ref{fig-universality}, in the time evolution of the many-particle position variance. 
Time evolution of $\frac{1}{N} \Delta^2_{\hat X}$ (a) when the initial condensed 
state is prepared in the left well and (b) when the initial condensed state is prepared in the right well. In each panel, the yellow curve represents the mean-field result while the color code for the MCTDHB results are explained in panel (c) and (d). Corresponding time evolution of 
$\frac{1}{N^2} \Delta^2_{\hat X}$ (c) when the initial BEC is prepared in the left well and (d) when the initial BEC is prepared in the right well.
Results are obtained by the MCTDHB method with $M=2$ orbitals. For further
details see the text. The quantities shown are dimensionless.}
\label{fig-universality-var}
\end{figure}

\section{Summary and concluding remarks}
\label{conclusions}

Summarizing, we have examined how the BJJ dynamics is affected by the loss of symmetry of the confining double well potential for different interaction $\Lambda$. In an asymmetric double well, the two wells are no longer equivalent. Therefore, we have studied the dynamics by preparing the condensate initially in both the left and right wells. We have analyzed the dynamics by examining the time evolution of three physical quantities  viz., the
survival probability, depletion or fragmentation, and, the many-particle position and momentum variances.

We find that the impact of the asymmetry of the trap depends on the interaction $\Lambda$ and the initial well. Overall, there is a 
suppression of tunneling between the two wells. However, the repulsive inter-atomic interaction facilitates the tunneling between the two wells when 
BEC is initially in the left well whereas the tunneling is further suppressed for starting the dynamics from the right well. For a sufficiently strong 
interaction $\Lambda$, the condensate becomes fragmented with time and the degree of fragmentation $f$ depends on the asymmetry $C$ and the initial well.  {The time evolution of the $\frac{1}{N} \Delta^2_{\hat X}$ bears prominent signature of the depletion of the system and deviates from its corresponding mean-field dynamics even for a weak $\Lambda$. In an asymmetric double well, both the frequencies and the amplitudes of the oscillations of $\frac{1}{N} \Delta^2_{\hat X}$ are found to be affected by the asymmetry. The dynamics of $\frac{1}{N} \Delta^2_{\hat X}$ in an asymmetric double well trap also bears signatures of the breathing-mode oscillations in addition to the density oscillations. However, the signatures of the breathing-mode oscillations are more prominent in the time evolution of the many-particle variance $\frac{1}{N} \Delta^2_{\hat P}$. While  in the time evolution of $\frac{1}{N} \Delta^2_{\hat P}$, breathing-mode oscillations are more prominent than the density oscillations for the symmetric double well, both are distinctly visible in case of the asymmetric double well. Since breathing-mode oscillations arise from coupling to higher energy bands, such features are beyond the scope of the Bose-Hubbard dimer.}    

An important observation of our study is the universal fragmentation dynamics of asymmetric BJJ. However, the degree of universal fragmentation 
for BECs consisting of different $N$ corresponding to the same $\Lambda$, depends on the initial well. {Universality of fragmentation is found to manifest in the same mean saturation value of the $\frac{1}{N^2} \Delta^2_{\hat X}$ for different $N$ corresponding to the same $\Lambda$ at the many-body level. {This means that the fluctuations of the positions of the particles in the junction show a universal behavior.} %However, this saturation value also depends on the initial well in conformity with the behavior of the universal fragmentation dynamics.
}

Macroscopic quantum tunneling is a fundamental quantum effect and is the underline mechanism for many physical events like Josephson junction. Also, BJJ is a paradigmatic device for understanding coherent quantum phenomena with potential applications in quantum interference technology, precision measurement, sensing, and, quantum metrology, etc. Particularly, in quantum interferometer, asymmetry of the trapping potential can be used as a means to shift the relative phase of the interferometer arms. {Also, in view of a growing area of quantum science and technology, there is a strong need for accurate many-body characterization of BJJs which is able to take into account all dominant and participating degrees of freedom.}
%Though a qualitative description may be good enough for understanding some of the underlying physics, an accurate quantitative description of the dynamics is vital for the technological applications of weak-link-based devices, such as double-slit interferometers.}
%Therefore, our study is relevant for offering a more realistic explanation to many quantum phenomena and also for finding new applications. 

\begin{acknowledgments}
This research was supported by the Israel Science Foundation (Grant No. 600/15). %We are grateful to Shachar Klaiman and Alexej Streltsov for discussions. 
Computation time on the High Performance Computing system Hive of the Faculty of Natural Sciences at University of Haifa and on the Cray XC40 
system Hazelhen at the High Performance Computing Center Stuttgart (HLRS) is gratefully acknowledged. SKH gratefully acknowledges 
the continuous hospitality at the Lewiner Institute for Theoretical Physics (LITP), Department of Physics, Technion - Israel Institute of Technology.
\end{acknowledgments}

\appendix*
%\section{Time evolution of the momentum variance of a weakly-interacting Bose gas}\label{J}
\section{{Further details of the numerical computations and their convergence}}\label{I}

{Here we discuss the details of our numerical computations. We remind that the ansatz in MCTDHB theory is taken as the superposition 
of all possible permanents constructed by distributing $N$ particles in $M$ time-dependent orbitals which are then determined by a time-dependent variational 
principle. Further for $M=1$, the ansatz Eq.~(\ref{MCTDHB_Psi}) boils down to the mean-field ansatz, and using the time-dependent variational method with
this ansatz gives the time-dependent Gross-Pitaevskii equation. 
Therefore, with our method, we can study the system at the mean-field level simply by considering $M=1$ orbital. On the other hand, using a finite number $M$ of orbitals, 
subject to the numerical convergence of the quantities of interest, we can get a numerically accurate many-body description of the system. 
Here we point out that in the limit $M \rightarrow \infty$, the set of permanents $\{\vert\vec{n};t\rangle\}$ spans the complete $N$-boson Hilbert space and 
thus the expansion Eq.~(\ref{MCTDHB_Psi}) is exact, but in numerical calculations, computational limitations rule out that option. At the same, time-dependence of the 
permanents as well as the expansion coefficients allows one to consider a much shorter expansion than if only the expansion coefficients are taken to be time-dependent
and thereby leads to a significant computational advantage. } 

%Figure 9
\begin{figure}[h]
\begin{center}
\begin{tabular}{cc}
\includegraphics[width=0.5\linewidth]{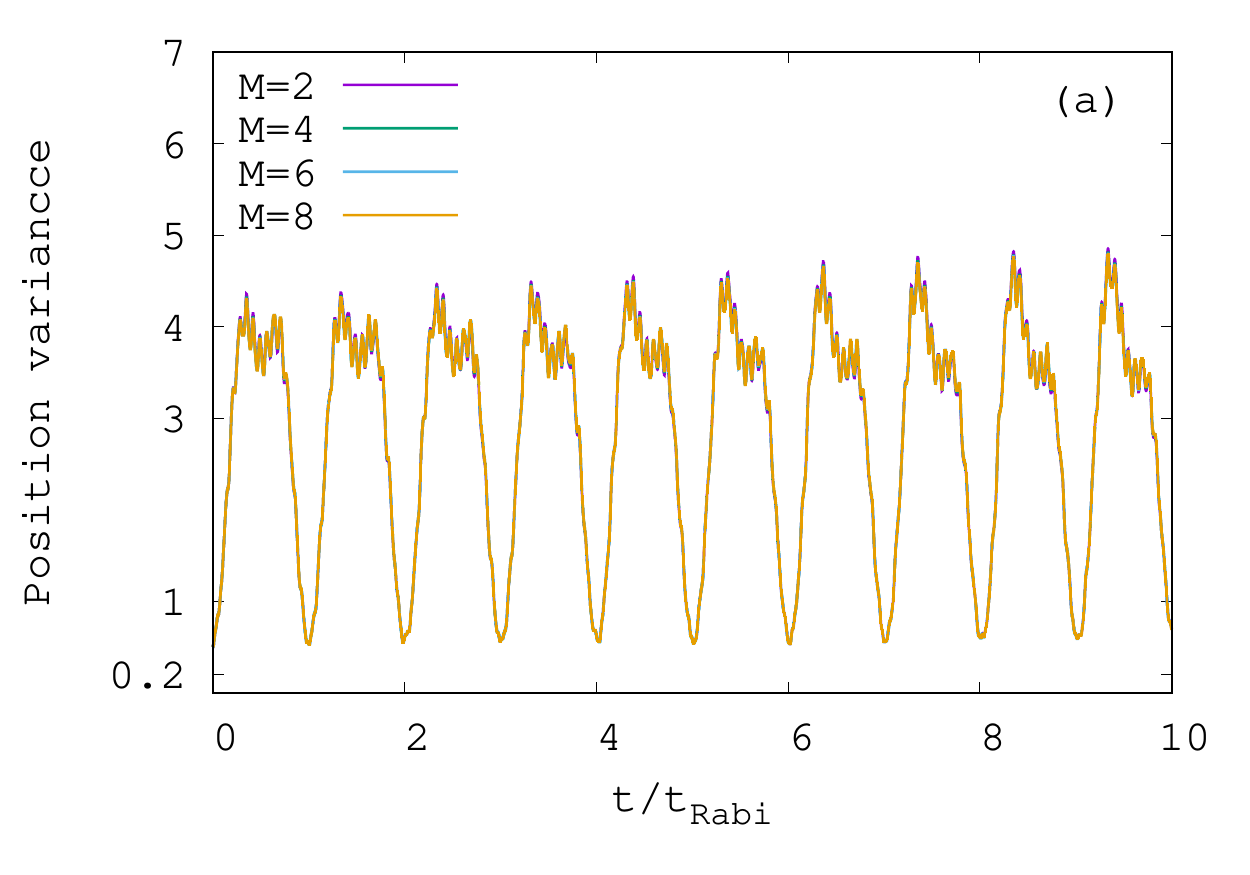} & 
\includegraphics[width=0.5\linewidth]{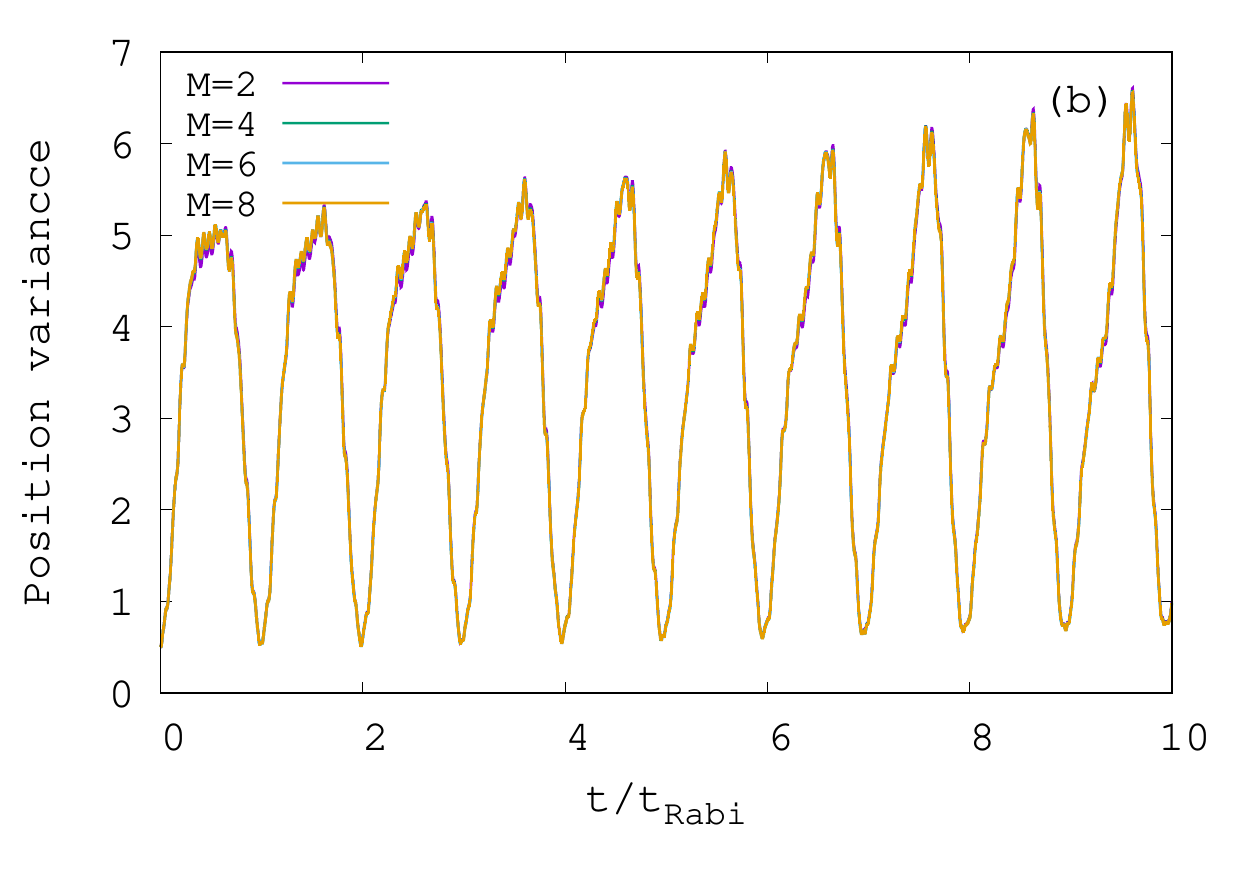} \\
\includegraphics[width=0.5\linewidth]{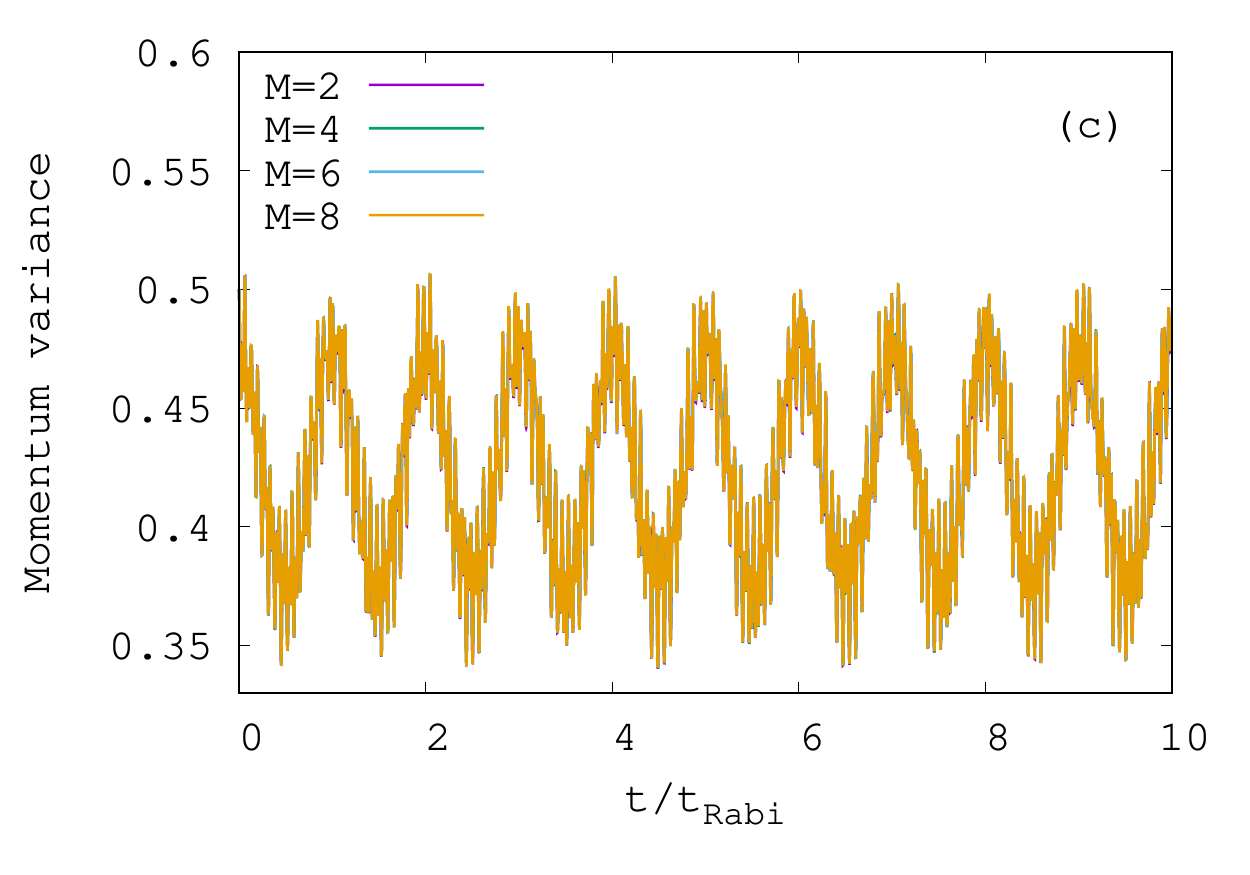} &
\includegraphics[width=0.5\linewidth]{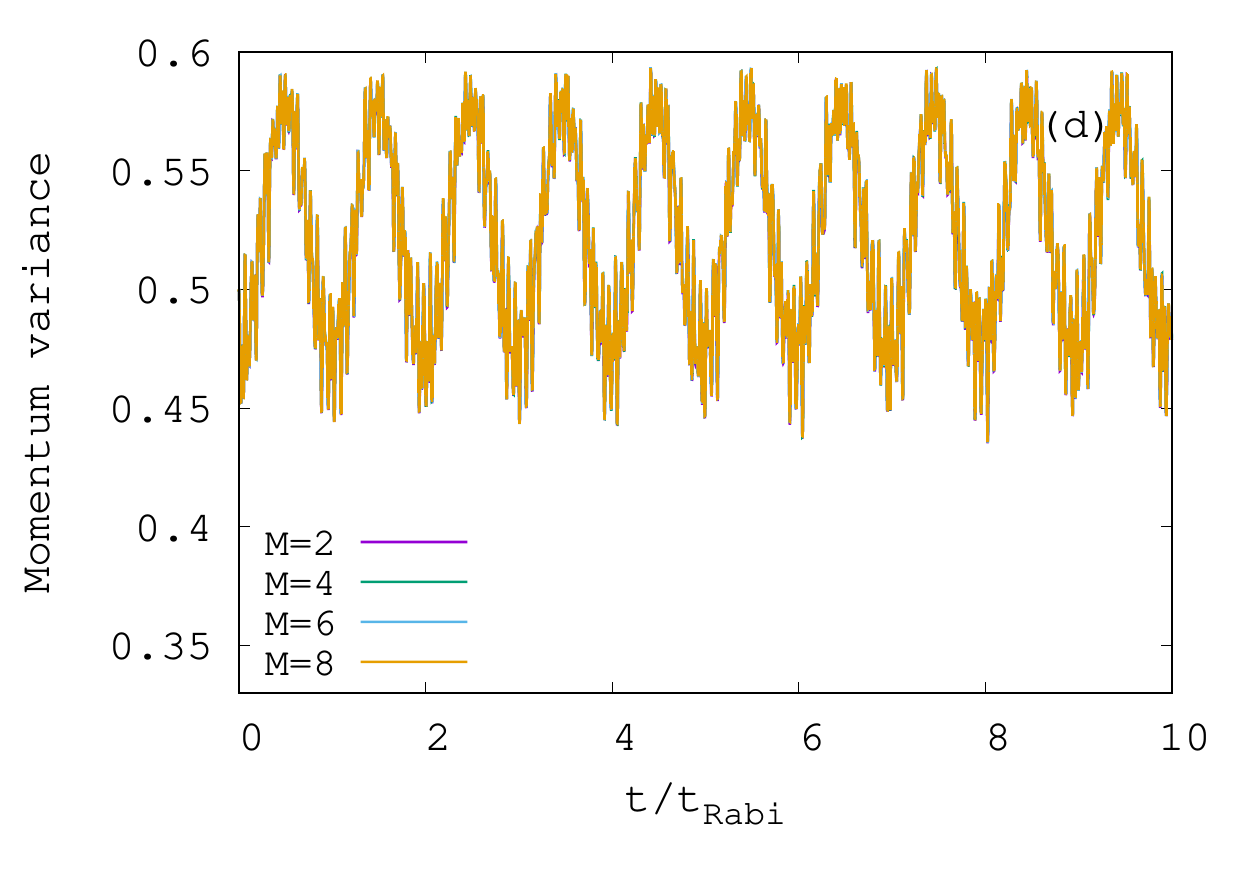} \\
\end{tabular}
\end{center}
\caption{Convergence of variances of the many-particle position and momentum operators with respect to the orbital number $M$ for a system of $N=10$ interacting bosons 
and $\Lambda=0.01$ in the asymmetric double well trap of asymmetry $C=0.01$. (a) Time-evolution of the many-particle position variance $\frac{1}{N} \Delta^2_{\hat X}$ computed by 
MCTDHB method with different $M$ when the initial condensate state is prepared in the left well. (b) The corresponding result for $\frac{1}{N} \Delta^2_{\hat X}$ when the initial 
condensed state is prepared in the right well. (c) Time evolution of the momentum variance $\frac{1}{N} \Delta^2_{\hat P}$ corresponding to (a). 
(d) Time evolution of the momentum variance 
$\frac{1}{N} \Delta^2_{\hat P}$ corresponding to (b). For details see text. Color codes are explained in each panel. The quantities shown are dimensionless.  }
\label{fig-convergence1}
\end{figure}

In our numerical {calculations}, {the many-body Hamiltonian is represented by 128 exponential discrete-variable-representation (DVR)
grid points (using a Fast Fourier transformation routine) in a box size [-10,10). We obtain the initial state for the time 
propagation, the many-body ground state of the BEC either in the left well or in the right well, 
by propagating the MCTDHB equations of motion {[Eq.~(\ref{MCTDHB1_equ})]} in imaginary time~\cite{MCHB, Lode2012}. For our numerical computations,
we use the numerical implementation in~\cite{Streltsov1,Streltsov2}.} 
We keep on {repeating the computation with} increasing $M$ until convergence is reached 
{and thereby obtain the numerically accurate results}.

{Below we demonstrate the numerical convergence of the many-particle position and 
momentum variances.} We already discussed in the text that the variance of any quantum operator is much more sensitive to the many-body effects compared to the oscillations in the 
survival probabilities and the fragmentation $f$. Actually, it is seen that the convergence of the momentum variance requires more numerical resources than the convergence of the 
position variance. Therefore, demonstration of convergence of the position and momentum variances will automatically {imply} 
the convergence of the survival probabilities and the fragmentation $f$ with respect to $M$.

{In an asymmetric double well, the two wells are not equivalent and therefore, in the text we discussed the dynamics of the system separately for 
preparing the initial BEC in the left well and in the right well. Accordingly here also, we discuss the convergence for both the cases separately, 
first when the initial BEC is prepared in the left well and then when the condensate is initially in the right well. We consider a system of $N=10$ interacting bosons 
in an asymmetric double well with the asymmetry $C=0.01$.}
Since in the limit $N \rightarrow \infty$, keeping $\Lambda$ fixed, the 
many-body effects diminish {and the density per particle of the system} converge to its corresponding mean-field values~\cite{Lieb_PRA,Lieb_PRL,Erdos_PRL,MATH_ERDOS_REF}, 
the convergence of the quantities for higher 
$N$ values considered in the text (but same $\Lambda$ values considered here) are actually better than what is shown below.

%Figure 10
\begin{figure}[h]
\begin{center}
\begin{tabular}{cc}
\includegraphics[width=0.5\linewidth]{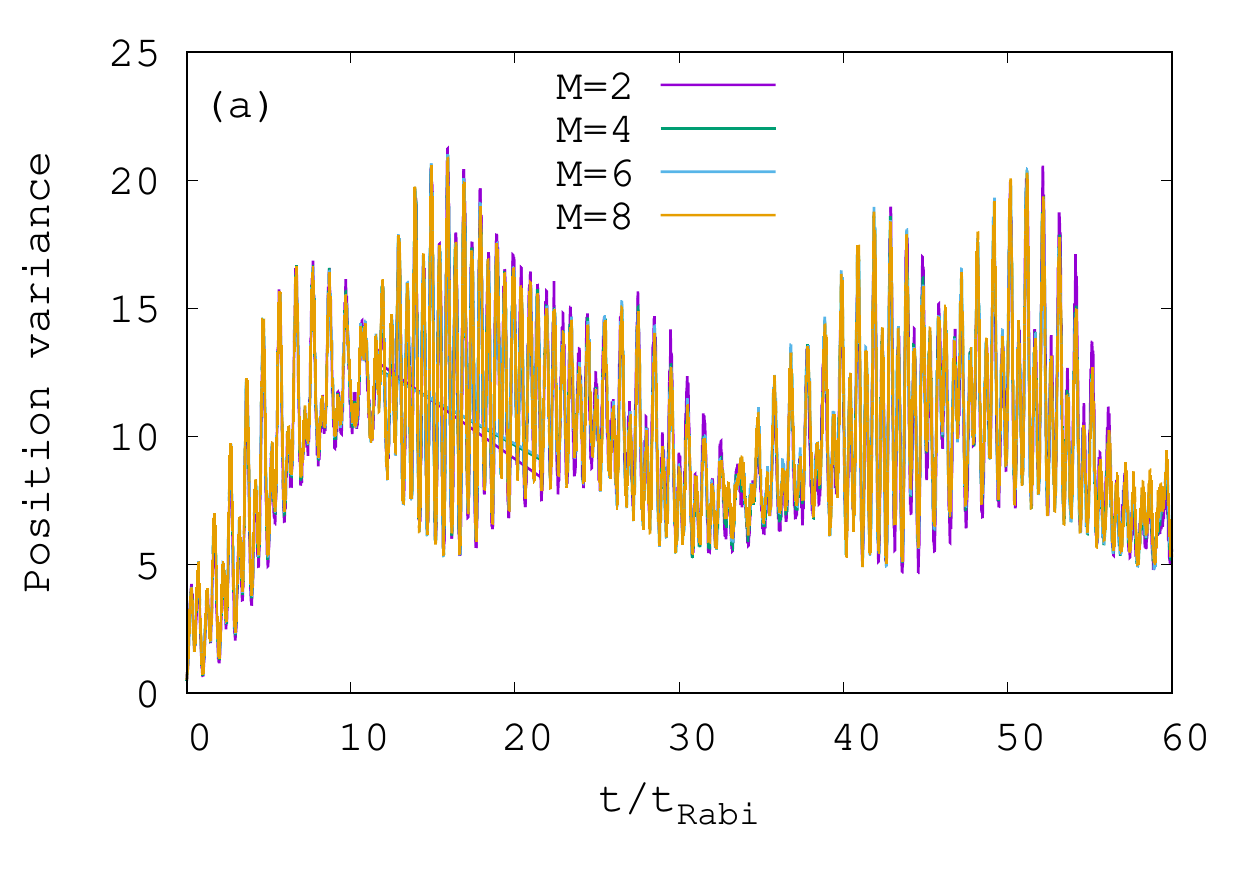} &
\includegraphics[width=0.5\linewidth]{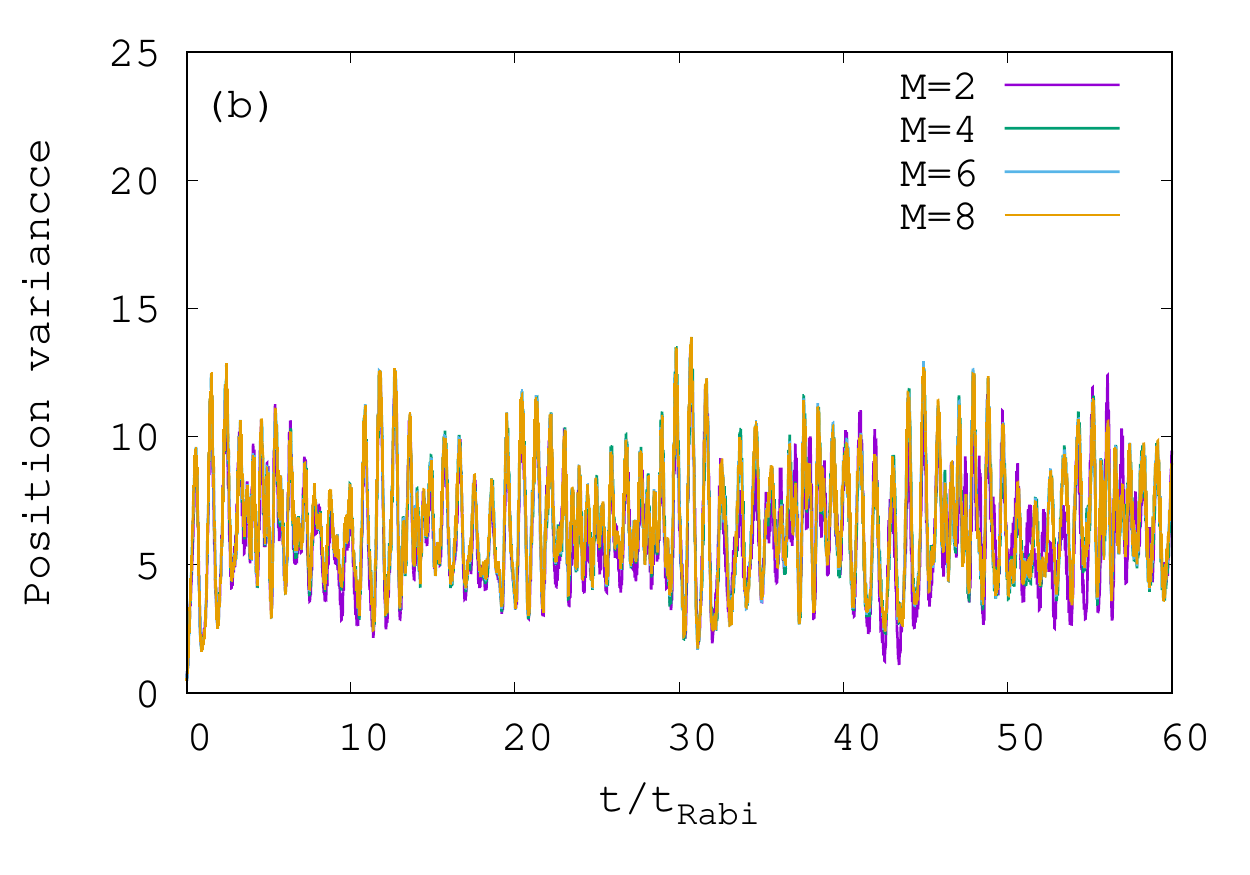}\\
\includegraphics[width=0.5\linewidth]{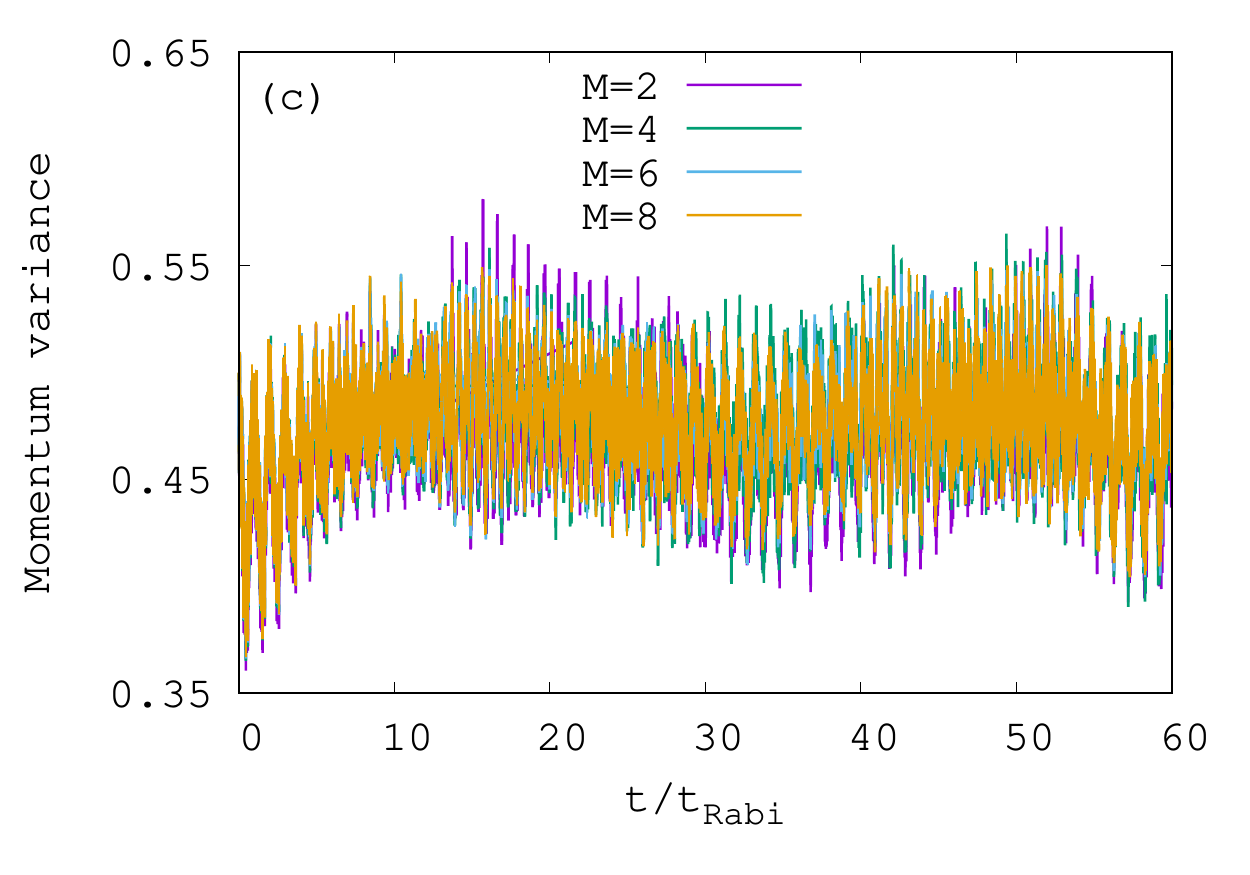} &
\includegraphics[width=0.5\linewidth]{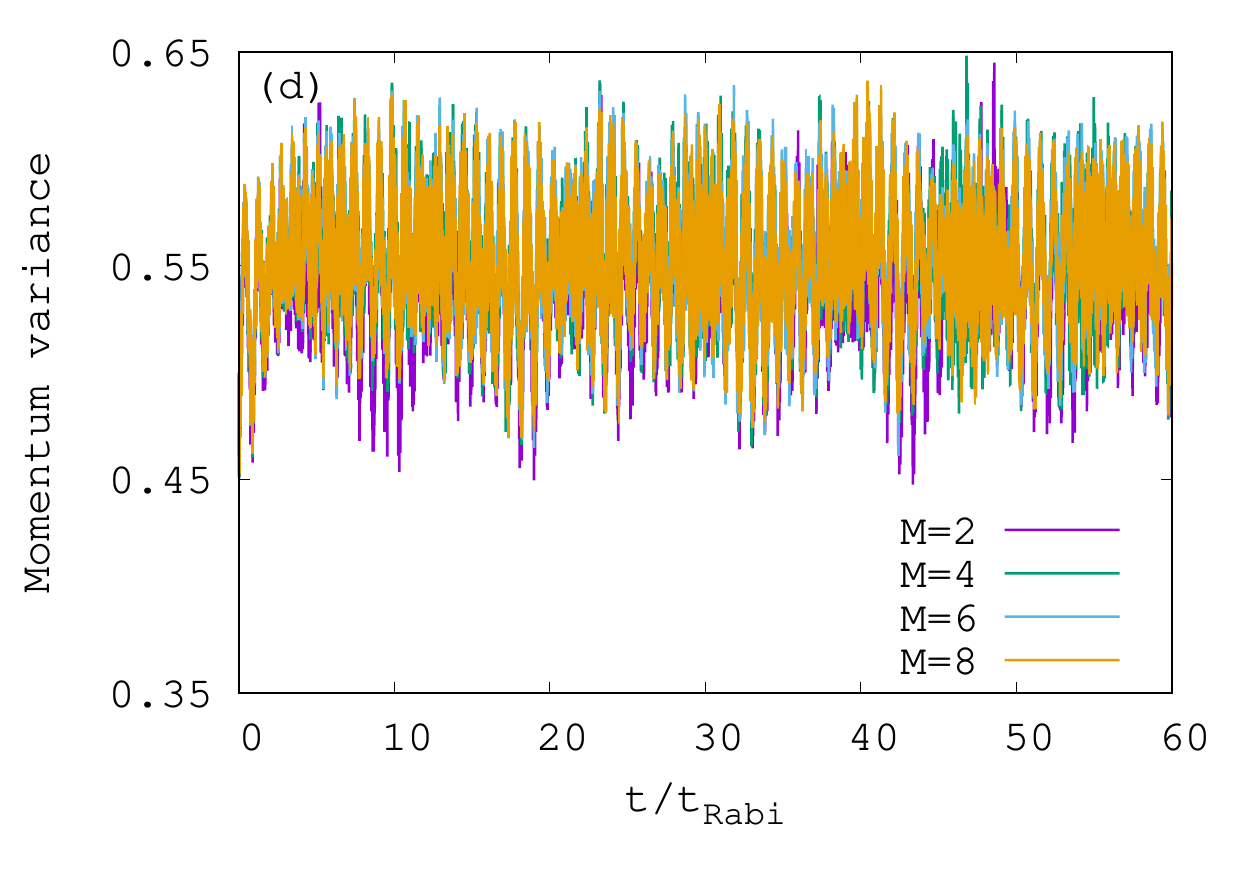}  \\
\end{tabular}
\end{center}
\caption{Convergence of variances of many-particle position and momentum operators with respect to the orbital number $M$ for a system of $N=10$ interacting bosons 
and $\Lambda=0.1$ in the asymmetric double well trap of asymmetry $C=0.01$. (a) Time-evolution of the many-particle position variance $\frac{1}{N} \Delta^2_{\hat X}$ computed by 
MCTDHB method with different $M$ when the initial condensate state is prepared in the left well. (b) The corresponding result for $\frac{1}{N} \Delta^2_{\hat X}$ when the initial 
condensed state is prepared in the right well. (c) Time evolution of momentum variance $\frac{1}{N} \Delta^2_{\hat P}$ corresponding to (a). (d) Time evolution of momentum variance 
$\frac{1}{N} \Delta^2_{\hat P}$ corresponding to (b). For details see text. Color codes are explained in each panel. The quantities shown are dimensionless.}
\label{fig-convergence2}
\end{figure}

We first consider $\Lambda=0.01$.
In Fig.~\ref{fig-convergence1} (a) and (b) we plot $\frac{1}{N} \Delta^2_{\hat X}$ computed with $M=2,4,6,$ and 
$8$ for starting the dynamics from the left and right well, respectively. We see that, as discussed in the main text, 
$\frac{1}{N} \Delta^2_{\hat X}$ exhibits a slow oscillatory growth for both wells. 
{Furthermore, we see that there is an overall oscillation of $\frac{1}{N} \Delta^2_{\hat X}$ in time with a frequency equal to twice the Rabi frequency. 
Also, on top of the peaks of these oscillations, there is another oscillation with a higher frequency but smaller amplitude. 
The origin of these oscillations are discussed in the main text, see Sec.~\ref{variance}. Here,
we observe that, for both cases, the results for $M=2,4,6,$ and $8$ are in very good agreement with each other, such that not only 
the overall oscillations of $\frac{1}{N} \Delta^2_{\hat X}$ due to the density oscillations but also the 
small amplitude high-frequency oscillations on top of the peaks of the first ones are accurately described with $M=2$.}

{To demonstrate the convergence of $\frac{1}{N} \Delta^2_{\hat P}$, in Fig.~\ref{fig-convergence1}(c) and (d), we plot the $\frac{1}{N} \Delta^2_{\hat P}$ 
computed with $M=2,4,6,$ and $8$ orbitals for preparing the condensate initially in the left and the right well, respectively. We clearly see that there are two oscillations 
associated with the time evolution of $\frac{1}{N} \Delta^2_{\hat P}$ on top of one another, one with a larger amplitude and frequency equal to twice the Rabi frequency and the 
other one with a smaller amplitude but higher frequency. Once again, we see that the results of $\frac{1}{N} \Delta^2_{\hat P}$ for different $M$ practically overlap with each other, 
and the $M=2$ is sufficient to describe both oscillations accurately.}

%Figure 11
\vglue +0.6 truecm
\begin{figure}[h]
\begin{center}
\begin{tabular}{cc}
\includegraphics[width=0.50\linewidth]{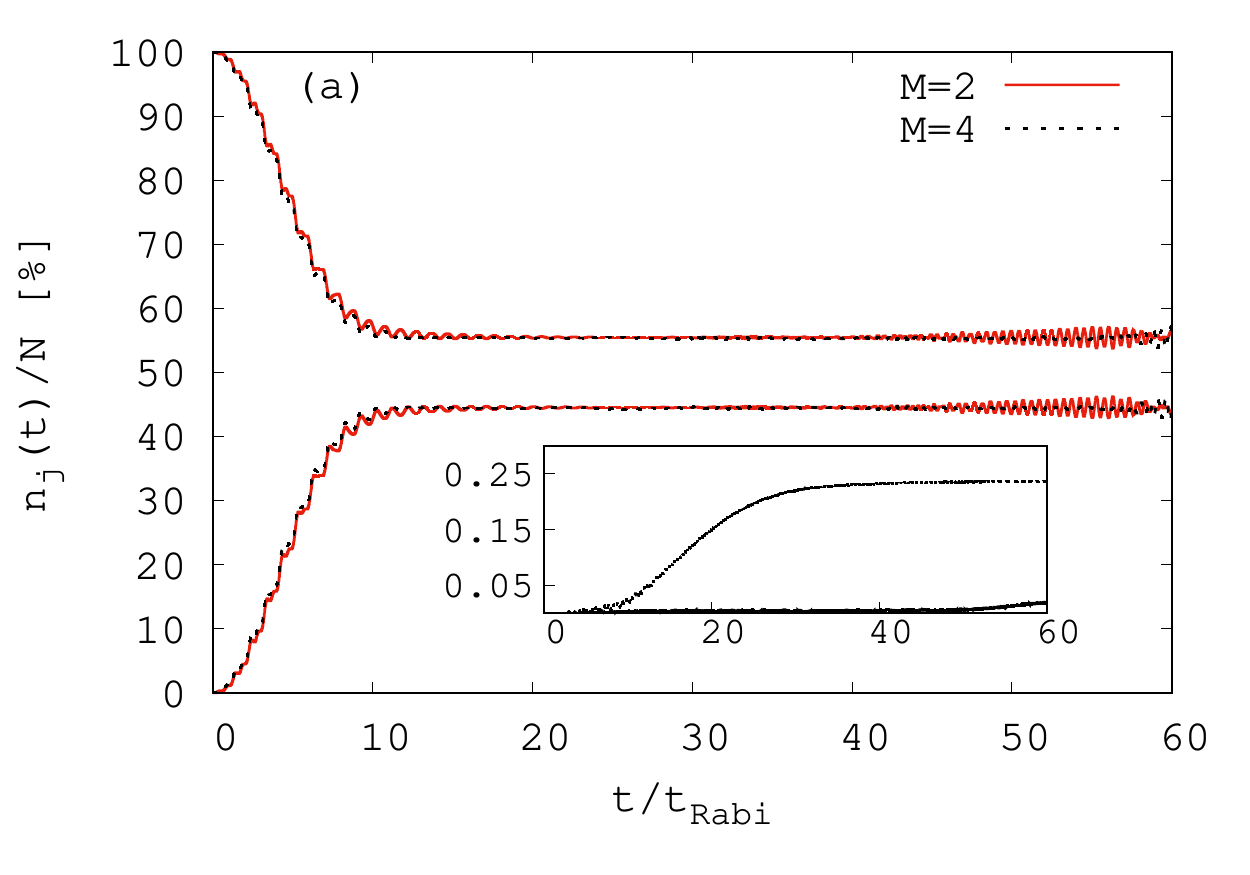} & 
\includegraphics[width=0.50\linewidth]{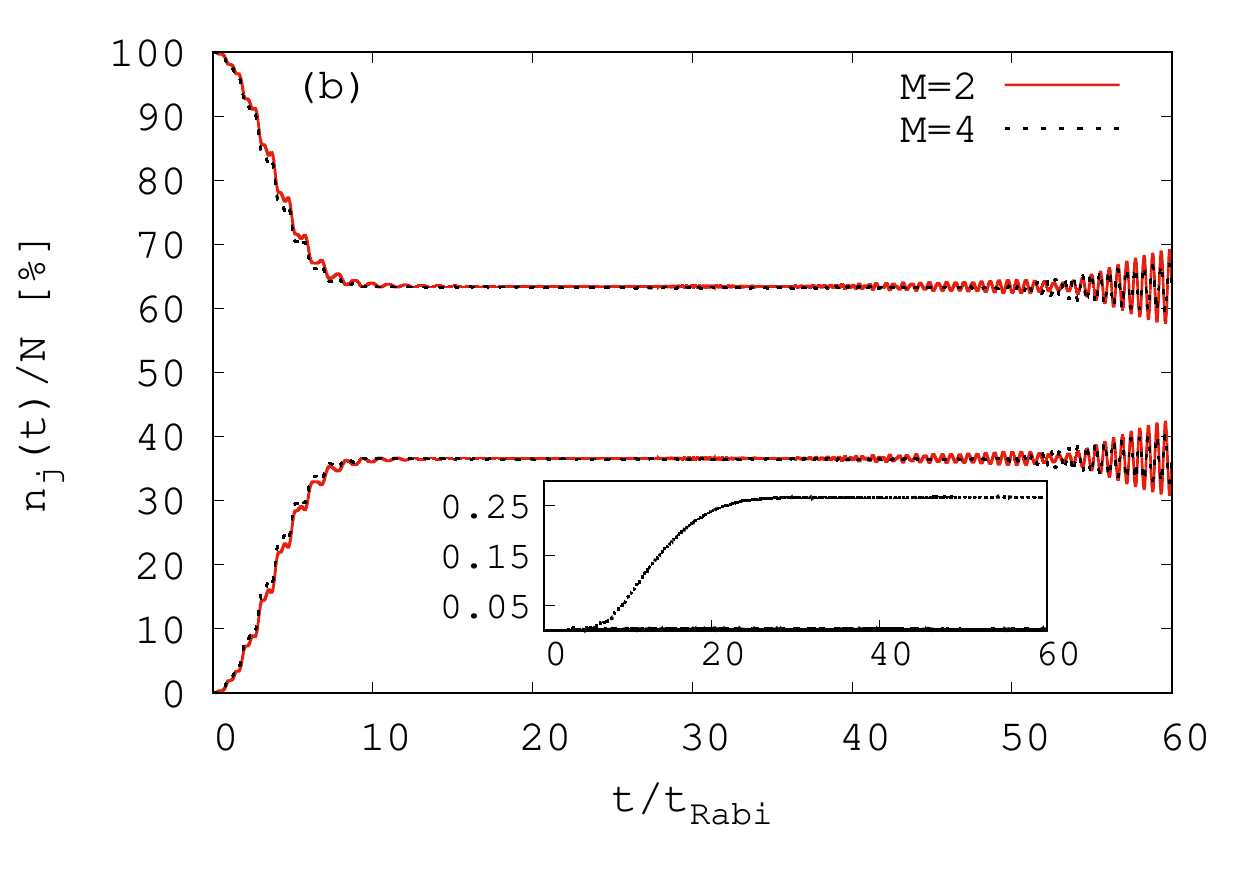}  \\
\end{tabular}
\end{center}
\caption{Convergence of the universal fragmentation dynamics with respect to the orbital number $M$ for a system of $N=100$ interacting bosons 
and $\Lambda=0.1$ in the asymmetric double well trap of asymmetry $C=0.001$. (a) Initially the condensate is prepared in the left (lower) well. 
(b) The condensate prepared in the right (higher) well prior to 
the dynamics in the asymmetric double well. In each panel, the first two natural occupations viz., $\frac{n_1}{N}$ (upper curve) and $\frac{n_2}{N}$ 
(lower curve) are shown for computations with $M=2$ and $M=4$ orbitals in the MCTDHB method. Color codes are explained in each panel.
In the insets of each panel, the higher natural occupations $\frac{n_3}{N}$ (upper curve) 
and $\frac{n_4}{N}$ (lower curve) computed by using $M=4$ orbitals in the MCTDHB method are shown. See text for further details. The quantities shown are dimensionless.}
\label{fig-convergence3}
\end{figure}

{Next, we consider the convergence for the stronger interaction $\Lambda=0.1$. 
We plot $\frac{1}{N} \Delta^2_{\hat X}$ computed with $M=2,4,6,$ and $8$ time-adaptive orbitals for starting the dynamics from the left and 
right wells in Fig.~\ref{fig-convergence2} (a) and (b), respectively. As discussed in Sec.~\ref{variance}, there is an equilibration-like effect following which 
$\frac{1}{N} \Delta^2_{\hat X}$ oscillates about a mean saturation value $\frac{1}{N} \Delta^2_{\hat X}\rvert_{sat}$. 
However, for $N=10$ the system size is very small and this equilibration is less pronounced.  
Clearly, the results of $\frac{1}{N} \Delta^2_{\hat X}$ 
obtained with $M=2$ accurately describe all features of the time evolution of $\frac{1}{N} \Delta^2_{\hat X}$ while those for higher $M$
practically overlap with each other. The corresponding results of $\frac{1}{N} \Delta^2_{\hat P}$ for starting the dynamics from the left and right wells, computed with $M=2,4,6,$
and $8$ orbitals, are shown in Fig.~\ref{fig-convergence2}(c) and (d), respectively. 
We point out that in the main text, $\frac{1}{N} \Delta^2_{\hat P}$ is not computed for this interaction strength.
For both cases, $\frac{1}{N} \Delta^2_{\hat P}$ keeps on oscillating about a 
mean value which depends on the well in which the condensate is initially  prepared. Again, we see that the computations with $M=2$ orbitals describe all the features of the time
evolution of $\frac{1}{N} \Delta^2_{\hat P}$ and, as $M$ increases, the curves corresponding to higher $M$ starts to practically overlap with each other.  }

Finally, to explicitly show that the convergences of other quantities are also achieved with the same $M$ as the variances, for the same system parameters, 
below as an example, we consider the convergence of the natural occupation numbers $\frac{n_j}{N}$. In Fig.~\ref{fig-convergence3} we plot $\frac{n_j}{N}$ for a system of $N=100$ 
and $\Lambda=0.1$ computed with $M=2$ and $4$ orbitals for preparing the initial condensate in the left [panel (a)] and right [panel (b)] wells of an asymmetric double well with 
$C=0.001$. We found that the computation with $M=4$ reproduces the same $f$ as with $M=2$ for both initial wells: The curves for the two largest occupation numbers
$\frac{n_1}{N}$ and $\frac{n_2}{N}$ almost completely overlap with the corresponding results obtained with $M=2$ and saturate about $\frac{n_1}{N} \approx 55\%$ and 
$\frac{n_2}{N} \approx 45\%$ for the left well and $\frac{n_1}{N} \approx 65\%$ and $\frac{n_1}{N} \approx 35\%$ for the right well, respectively. 
Also, the two smaller occupation numbers, viz. $\frac{n_3}{N}$ 
and $\frac{n_4}{N}$ obtained with $M=4$ are negligibly small for both cases. While $\frac{n_3}{N}$ grows slowly with time only to saturate around $0.25\%$, 
$\frac{n_4}{N}$ shows very little increment from its initial near-zero value. Thus the convergences for quantities like $\frac{n_j}{N}$ are achieved with the same $M$ as the variances
for the same system parameters. Moreover, the near perfect agreement between the two results obtained with $M=2$ and $M=4$, respectively, along with the negligibly small values of 
$\frac{n_3}{N}$ and $\frac{n_4}{N}$ show that convergences improve with $N$ keeping $\Lambda$ fixed for a repulsive interaction. 
Finally and importantly, it also demonstrates that the universality of fragmentation is
a robust many-body phenomenon and does not fissile out by using  larger numbers $M$ of orbitals. 
%\section*{References}

%\bibliography{manuscript}
%\end{document}

\end{document}